Banner appropriate to article type will appear here in typeset article

# Bubble coalescence dynamics in a high-Reynolds number decaying turbulent flow

**Vivek Kumar, Prasoon Suchandra, Ardalan Javadi, Suhas S. Jain, and Cyrus Aidun**
George W. Woodruff School of Mechanical Engineering, Georgia Institute of Technology, Atlanta, GA 30332, USA
Renewable Bioproducts Institute, Georgia Institute of Technology, Atlanta, GA 30332, USA
**Corresponding author:** Cyrus Aidun, cyrus.aidun@me.gatech.edu

This study experimentally investigates bubble size evolution and void fraction redistribution in an unexplored, coalescence-dominated regime of decaying turbulent bubbly flow. The flow is generated downstream of a regenerative pump in a duct, with bulk Reynolds number (Re) $\sim O(10^5)$, Taylor-scale Reynolds number (Re$_\lambda$) $\sim O(10^3)$ and void fraction ($\phi$) $\sim O(1\%)$, where inlet turbulence is extremely intense (turbulent intensity > 30%) but decays rapidly along the duct. Shadowgraph imaging and particle shadow velocimetry are used for measurements. The measured turbulent dissipation in the duct flow decays as $\varepsilon \sim \mathcal{L}^{-2}$, where $\mathcal{L}$ is the axial position, in close agreement with the homogeneous–isotropic turbulence prediction of $\varepsilon \sim \mathcal{L}^{-2.2}$. High-speed imaging and statistical analysis reveal that bubble coalescence dominates over breakup across most of the domain, leading to monotonic growth in the Sauter mean diameter ($d_{32}$) and progressive broadening of the bubble size distribution. The normalised extreme-to-mean diameter ratio ($\mathcal{D}$) increases axially and asymptotically saturates at $\sim 2.2$, indicating the emergence of a quasi-self-similar bubble size distribution. The probability density function of bubble diameter exhibits a dual power law tail with exponents $-10/3$ and $-3/2$ near the duct inlet, where the flow is coalescence-dominated. However, after a few hydraulic diameters, a single $-3/2$ power law scaling emerges, indicating a regime of pure coalescence in which all bubbles are smaller than the Hinze scale. The cumulative distribution with a $d/d_{32}$ exponent ($\sim 1.3$) emerges only after the size distribution stabilises. Although classical Hinze scaling gives $d_H \propto \mathcal{L}^{0.9}$, our theory for $d_{32}$ and $d_{99.8}$ (99.8th percentile bubble diameter) in a pure-coalescence regime predicts the slower law $\propto \mathcal{L}^{0.5}$, which our experimental results confirm – indicating negligible breakup and sub-Hinze growth. Concurrently, in contrast to current models, transient void fraction ($\phi$) profiles evolve from nearly uniform to sharply core-peaked Gaussian distributions in the developing regime, with increasing centerline values and decreasing near-wall values, due to lift-force reversal. These results provide the first spatially resolved characterization of coalescence-dominated bubbly flows at high Re, advancing the design of industrial systems as in nuclear cooling and multiphase forming processes (e.g., paper manufacturing, chemical reactors).

## 1. Introduction

Multiphase turbulent flows, where gases and liquids mix, are ubiquitous in industrial processes and natural phenomena. In applications ranging from chemical reactors and oil-gas pipelines (Delnoij et al. 1997) to oceanic wave breaking (Deane & Stokes 2002a), the interaction of bubbles with turbulence critically affects mass transfer, reaction rates (Clift et al. 2005), and flow behavior. For example, the coalescence and breakup of bubbles in turbulence governs interfacial area and influences efficiency in systems like emulsifiers, heat exchangers, and wastewater aeration tanks (Martínez-Bazán et al. 1999; Coulaloglou & Tavlarides 1977). High Reynolds number [Re= $(\rho V D_i)/\mu$), where $V$ is the impeller tip speed and $D_i$ is the impeller diameter] bubbly flows are especially important in devices such as multiphase pumps, where intense turbulence is deliberately generated to disperse gas into liquid (Javadi et al. 2025). In such pump operations, bubbles can significantly alter performance: their formation, coalescence, and breakup can enhance mixing, but also cause loss of efficiency if not controlled (Kolmogorov 1949; Hinze 1955a; Javadi et al. 2025). For instance, in a multiphase pump impeller, strong turbulence fragments bubbles into smaller bubbles and influences the downstream two-phase flow (Zhang et al. 2018). Given the prevalence of bubbles in turbulence in engineering (e.g., bubble columns, pipelines, nuclear coolant loops) and in natural phenomena (e.g., breaking waves, volcanic eruptions), understanding coalescence/breakup dynamics under extreme turbulence is both fundamentally and practically important (Liao & Lucas 2009, 2010a).

The bubble size distribution mainly depends on two factors: coalescence and breakup in turbulent flow. These mechanisms have been extensively studied theoretically and computationally in turbulent environments, where energy dissipation, inertial forces, and interfacial dynamics dictate the evolution of the bubble size distribution (Li & Liao 2024; Li et al. 2024). Bubble breakup in turbulence arises when inertial stresses exceed surface tension, fragmenting bubbles above a critical Weber number (Kolmogorov 1949; Hinze 1955a; Ni 2024). Experiments by Martinez-Bazan et al. (1999) showed a two-regime breakup frequency: rising near the critical size and decreasing for very large bubbles. Na et al. (2022) reported a turbulence-driven power law size distribution in oceanic flows. Direct numerical simulation (DNS) studies revealed further detail: Vela-Martín & Avila (2022) demonstrated memoryless breakup even for sub-Hinze droplets, while Calado & Balaras (2024) highlighted enhanced deformation when bubble size matches energy-containing eddies. Models by Prince & Blanch (1990a), the energy cascade framework by D. K. R. Nambiar & Gandhi (1990), and Li & Liao (2024)'s turbulence modulated breakup relate breakup rates to turbulent dissipation and deformation. High Reynolds numbers amplify breakup via increased shear and energy cascade to small scales (Chen et al. 2021; Sajjadi et al. 2013), making the Weber number central to predicting fragmentation (Li & Liao 2024), with bulk-phase dissipation further modulating breakup dynamics (Nguyen et al. 2013).

Bubble coalescence in turbulent flows is governed primarily by collision frequency and coalescence efficiency (Scarbolo et al. 2015). Early foundational models, such as Coulaloglou & Tavlarides (1977), introduced population balance frameworks based on turbulent eddy collisions and liquid-film drainage mechanisms, establishing a fundamental framework for modeling bubble coalescence. Prince & Blanch (1990a) and Guo et al. (2016) further developed this approach by distinguishing between collision frequency and coalescence efficiency, explicitly incorporating turbulence, buoyancy, and viscous shear effects into their coalescence kernel. The Luo & Svendsen (1996) model refines these approaches by explicitly incorporating turbulence dissipation scales into predictions of collision and coalescence efficiency through film drainage modeling, proving especially effective at high void fractions (Luo & Svendsen 1996; D. K. R. Nambiar & Gandhi 1990). The recently developed Interfacial Area Transport Equation (IATE) further categorizes coalescence events into random collisions, wake-entrainment events, and hydrodynamic instabilities, offering a comprehensive framework widely accepted for detailed analyses of bubble interactions (Chen et al. 2021). Alternative approaches include energy-based models, such as Sovova's energy model, which postulates coalescence when the kinetic energy of colliding bubbles surpasses their surface energy (Liao & Lucas 2010a). Liao & Lucas (2010a) expanded this concept, advocating its direct applicability under turbulent conditions to emphasize the kinetic energy and surface energy balance during collisions.

In highly turbulent flows (e.g., Re $> 10^5$), the bubble collision frequency increases markedly as elevated turbulent kinetic energy and fluctuating eddies generate rapid relative motion and frequent encounters between bubbles (Lance & Bataille 1991; Serizawa et al. 1975). However, high shear forces and fluctuating velocities shorten the contact time during collisions, reducing coalescence efficiency by impeding liquid-film drainage and stabilizing the intervening film (Liao & Lucas 2010a; Hagesaether et al. 2000). Wake entrainment effects, in which larger bubbles induce secondary flow structures, further influence local bubble distributions and enhance collision heterogeneity. As a result, turbulence simultaneously promotes collisions while suppressing net coalescence, creating a regime in which frequent impacts do not necessarily lead to merging (Lance & Bataille 1991; Liao & Lucas 2010a; Hagesaether et al. 2000; Laakkonen et al. 2007a). These phenomena highlight the need to refine traditional coalescence models to account for the dual role of turbulence and the complex interplay between collision frequency, coalescence efficiency, and flow structures in high-Reynolds-number multiphase flows (Laakkonen et al. 2007a).

Empirical validation and modeling efforts have integrated these coalescence kernels into computational fluid dynamics (CFD) via population balance approaches. Comparisons of various breakup and coalescence model combinations in bubbly-flow simulations, such as bubble columns and stirred reactors, have demonstrated considerable variability in model predictions (Matiazzo et al. 2020). For instance, Laakkonen et al. (2005, 2006, 2007b) validated several models against experimental data from moderately turbulent stirred vessels, employing Multi-Size Group modeling (MUSIG) approaches to effectively reproduce bubble size distributions (Kamp et al. 2001; Matiazzo et al. 2020).

Experimental evidence also highlights coalescence dominance under less vigorous fully developed turbulence conditions. Razzaque et al. (2003b) studied air-water flows in horizontal pipes at moderate velocities (Re $\approx 10^5$), observing predominantly bubble coalescence into log-normal size distributions with stable downstream bubble spectra characterized by consistent $d_{99.8}/d_{32}$ ratios ($\sim 2.2$), where $d_{99.8}$ is the 99.8th percentile bubble diameter and $d_{32}$ is the Sauter mean diameter. These observations reinforce the critical influence of turbulence intensity on bubble population dynamics, underscoring the delicate interplay between coalescence and breakup mechanisms in turbulent multiphase flows.

Despite extensive research, significant gaps remain in our understanding of bubble dynamics in extreme and rapidly evolving turbulence, particularly at very high Reynolds numbers (Re $> 10^5$) immediately downstream of pumps or injectors. Such regions exhibit intense, spatially non-homogeneous turbulence, often reaching turbulence intensities around 50% far exceeding typical fully developed pipe flows (Javadi et al. 2025). Although critical in various industrial applications, bubble interactions in such intense and rapidly decaying turbulent fields remain inadequately studied. Existing theoretical frameworks, including the Interfacial Area Transport Equation (IATE) and Multi-Size Group (MUSIG) population balance models rely heavily on empirical closures calibrated under moderate or steady-state turbulence conditions (Kamp et al. 2001). These models fail to capture the pronounced spatial gradients in turbulence intensity and rapidly evolving bubble size distributions in developing turbulent flows, making their predictive reliability in intense turbulence downstream of pumps, ejectors, or mixers highly uncertain. Experimental data under such extreme turbulent conditions are notably scarce. Prior research predominantly addresses milder turbulence, resulting in classical log-normal bubble size distributions, or utilizes idealized conditions like homogeneous grid turbulence or confined bubble swarms (Bouche et al. 2012, 2014). Bouche et al. (2012, 2014) suggest that bubble size distributions significantly deviate from classical trends under extreme turbulence, necessitating systematic experimental validation to reconcile these deviations with existing theoretical models. A further critical knowledge gap involves the transient evolution of radial void fraction profiles under high-intensity turbulence. While literature confirms wall-peaked void fraction distributions for small bubbles at low void fractions due to lift forces (Serizawa et al. 1975; Shawkat et al. 2008; Hosokawa & Tomiyama 2009), the dynamic evolution from initial uniform distribution at high inlet turbulence levels in developing flow has not been systematically investigated.

The present study provides a first-of-its-kind experimental insight into bubble dynamics in a square duct under extreme pump-driven turbulence (bulk $Re > 10^5$ and $Re_\lambda \approx 900$). In this unique flow configuration, very high inlet turbulence intensity ($I \gtrsim 30\%$) fragments incoming bubbles to sizes near the classical Hinze scale, in equilibrium with the effect of breakup and coalescence (Hinze 1955b; Martinez-Bazan et al. 1999). As the flow convects downstream, turbulence decays rapidly along the axial direction. The Hinze scale ($d_H \propto \varepsilon^{-2/5}$, where $\varepsilon$ is turbulent dissipation) thus grows faster than the actual bubble diameters based on rate of coalescence ($d_{32}$, $d_{max} \propto \varepsilon^{1/3}$) (Prince & Blanch 1990a). This mismatch drives a coalescence-dominated regime: with very high rate of bubble collision frequencies as $\varepsilon$ still remains high, and net merging proceeds with negligible breakup. Such an extreme coalescence regime in a decaying turbulent flow has not been previously quantified (Ruth et al. 2022). The bubble-size evolves from the initial violent fragmentation at the pump outlet to downstream coalescence-driven growth as the turbulence subsides by over 90%. Bubble-size distribution exhibits a dual power law signature: a $-3/2$ slope during the coalescence-dominated regime (typical of *sub-Hinze* merging (Deane & Stokes 2002a; Ruth et al. 2022; Mostert et al. 2022)) and a $-10/3$ slope once bubble sizes exceed the evolving Hinze scale threshold (Martinez-Bazan et al. 1999; Ruth et al. 2022). Notably, buoyancy-driven segregation is suppressed due to the high bulk flow velocity, small bubble sizes, and moderate duct aspect ratio that minimize buoyancy effects (Drew & Lahey 1982; Clift et al. 2005; Magnaudet & Eames 2000). As a result, bubble behaviour remains effectively identical in vertical and horizontal ducts, broadening the applicability of our findings. In turbulent shear flows, inertial lift forces are known to dominate bubble lateral migration (Mazzitelli et al. 2003), further reinforcing orientation invariance. Finally, we quantify how void fraction and bulk liquid velocity modulate bubble dynamics: increasing $\phi$ raises collision frequency and accelerates coalescence, whereas higher bulk velocity sustains breakup and delays coalescence. The radial void-fraction profile evolves from nearly uniform near the inlet to sharply peaked Gaussian downstream, reflecting lift-induced migration (Mazzitelli et al. 2003).

 

The paper proceeds as follows. Section 2 describes the flow loop, design of experiments, control variables, and operating conditions (gas volume fraction, bulk velocity). Section 3 outlines the measurement methods - shadowgraph imaging for bubble detection/tracking and particle shadow velocimetry (PSV) for the turbulent field—and the computation of bubble statistics. Section 4 reports axial/radial trends in bubble-size distributions and turbulence, the shift from breakup- to coalescence-dominated regimes, and links to turbulence decay and dissipation, with comparisons to theory. Section 5 summarizes key findings and implications for modeling extreme-turbulence multiphase flows. Experimental conditions are compiled in Figure 2(a); symbols and acronyms appear in Table 4.

## 2. Experimental setup and experimental procedure

A schematic of the experimental setup is presented in Figure 1. The flow loop begins at a 100-gallon conical stainless-steel tank (Diboshi, 100 Gallon Conical Fermenter) filled with tap water. The tank connects via a flexible tubing (internal diameter: 17.8±0.02 mm, length ≈1 m) to the suction side of Pump I, a centrifugal pump (Dayton, PPLTAF23TDEG), whose speed is controlled by a variable frequency drive (VFD; Invertek Drives, ODE-3-320153-1042). The outlet of Pump I is connected to a Coriolis mass flow meter (Krohne MFM 4085K Corimass, type 300G+), shown as flow+pressure sensor, via a 1 m segment of the same 17.8±0.02 mm tubing. Downstream, the flow enters a short polycarbonate circular duct (ID 17.8±0.02 mm, length 15.24±0.03 cm) divided into two sections and connected via a 1 inch (2.54 cm) NPT T-junction. An inline stainless-steel sparger of $\frac{1}{4}$ inch (6.35 mm) NPT (2308-A04(00)-06-A00-GAS-AB, Moto Corp.) is installed in the direction of the flow, enabling aligned air injection. Compressed air is supplied through this sparger from a high-pressure compressor (maintained at 8 bar), regulated by a calibrated mass flow controller (Omega, FMA-A2309) placed in the supply line to regulate the injection rate precisely. This air–water mixing region is designed to align air injection with the bulk liquid flow and ensure fine initial dispersion.

Following the circular mixing duct, the flow enters Pump II, a regenerative turbine pump (MTH Pumps, Model E51M SS), via another segment of 17.8±0.02 mm tubing. Pump II is independently controlled using a dedicated VFD to maintain desired flow rates by adjusting motor frequency. To ensure the validity and independence of the setup, experiments were repeated with three different impellers, and the results were consistent with those reported here. The discharge of Pump II feeds into the test section - a vertical, square cross-sectional duct of internal dimensions 13.97±0.03 mm × 13.97±0.03 mm and a height of 60.96±0.30 cm (aspect ratio 40). This duct is constructed from a rigid polycarbonate frame sandwiched between two optically clear 95% transparent polycarbonate panels, allowing full optical access along the axial length for flow visualization and measurement. Bubble-size measurements were performed at multiple axial positions (x/D = $\mathcal{L}$ = 3.6 to 40, D = 13.97 mm) and four radial locations (centerline, near-wall, and intermediates) using shadowgraph-based imaging. A high-speed camera (Photron FASTCAM Nova R2) equipped with a high-magnification lens (Navitar Resolv4K) and synchronized with a halogen backlight (OSL2, Thorlabs) was used to capture the bubble field. The camera was interfaced with a data acquisition system to record high-resolution images at each measurement location. Particle shadow velocimetry (PSV) (Estevadeordal & Goss 2005; Khodaparast et al. 2013; Hessenkemper & Ziegenhein 2018; Jassal et al. 2025) was employed in parallel using similar visualization window to estimate turbulence statistics, including velocity fields, turbulent kinetic energy (*tke*, $k$), turbulence intensity ($\mathcal{I}$), and dissipation rate ($\varepsilon$) .

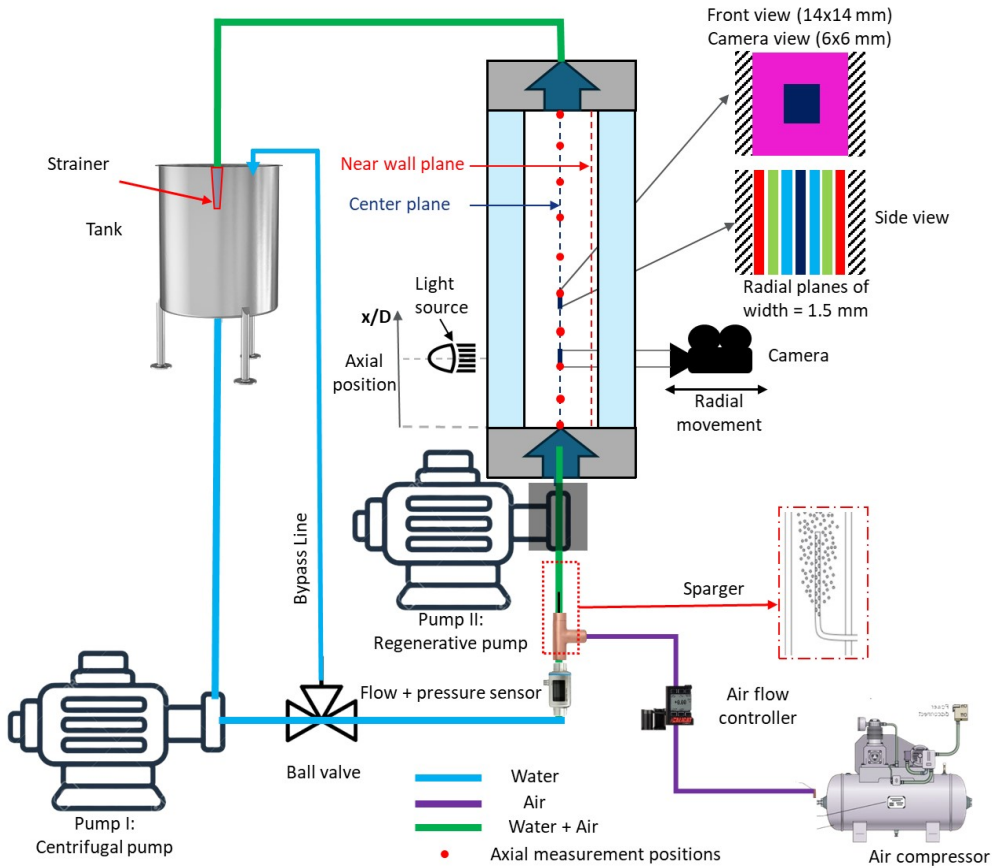

Figure 1. Schematic of experimental setup and flow loop. Red bubblet: ten axial positions. Front view: channel cross-section 14 × 14 mm (magenta); visualisation window 6 × 6 mm (dark blue). Side view: measurements in seven radial planes from near-wall (red) to center plane (dark blue), each 1.5 mm wide.

The outlet of the test section is connected to a return line that re-circulates the two-phase mixture back into the storage tank. A fine mesh strainer (4856K221, McMaster-Carr) is placed at the return to prevent large air slugs from re-entering the suction line. To ensure minimal air entrainment into Pump I, the return line terminates with a vertically mounted perforated bottle within the tank, which facilitates the complete separation of entrained air from the returning water. The tank's large cross-sectional area further aids in the natural escape of bubbles, while ensuring a bubble-free inlet condition at the pump suction. The absence of air entrainment was verified by capturing high-speed images under no-air-injection conditions. Thermal effects were minimal throughout the experiments. The only source of heat was viscous dissipation due to wall shear. The initial water temperature was maintained at 20 °C and continuously monitored. If the temperature deviated from 20±2 °C during extended operation, the water was drained and replaced to ensure thermal consistency across all measurements.

The experiments were conducted to investigate bubble dynamics in a highly turbulent, decaying duct flow over a broad range of operating conditions, as shown in Figure 2(a). Three bulk velocities, corresponding to volumetric flow rates of 72, 87, and 98 LPM (6.1–8.4 m/s), yielded bulk Re of $9.4 \times 10^4$, $1.14 \times 10^5$, and $1.29 \times 10^5$. For each velocity, three inlet void fractions ($\phi = 0.5, 1, 2\%$) were varied. Measurements were performed at ten axial positions spanning x= 0 − 40D, capturing the progressive evolution of the two-phase flow as turbulence decayed downstream. The corresponding Taylor-scale Re near the inlet ranged from 583 to 918. Radial variations were resolved at four discrete interrogation windows as seen in Figure 2(b).

| $\mathcal{L}$ | $Q$ $(V)$ LPM (m/s) | $\phi$ (%) | Re | Re$_\lambda(\mathcal{L} \approx 0)$ |
|---|---|---|---|---|
| 0, 3.6, 8.2, 12.7, 17.8, 21.8, 26.4, 30.9, 35.5, 40.0 | 72 (6.1) | 0.5 1 2 | 94.5K | 583 |
| | 87 (7.4) | 0.5 1 2 | 114.2K | 825 |
| | 98 (8.4) | 0.5 1 2 | 128.6K | 918 |

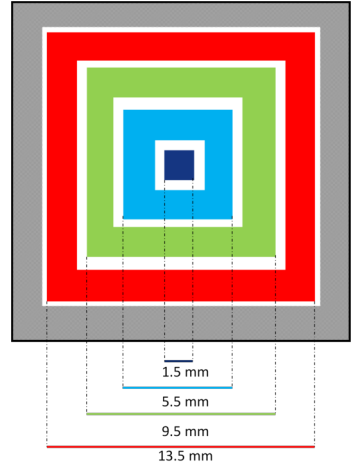

Figure 2. **Left:** Experimental conditions, including axial positions ($\mathcal{L}$), flow rates ($Q$), bulk velocity ($V$), void fraction ($\phi$), bulk Reynolds number (Re = $VD/\nu$), and Taylor Reynolds number (Re$_\lambda = \mathcal{U}\lambda/\nu$), where $D$ is the duct hydraulic diameter, $\nu$ is the liquid kinematic viscosity, $\mathcal{U}$ is the root-mean-square (rms) value of the velocity fluctuations, and $\lambda$ is the Taylor microscale. **Right:** Radial measurement locations in a $14 \times 14$ mm duct region. Each measurement plane [red (near wall), green, sky blue, dark blue (center)] is 1.5 mm thick and separated by 0.5 mm white spacing, and wall to red plane clearance is 0.25 mm.

## 3. Measurement techniques

### 3.1. *Shadowgraph for bubble imaging*

A high-speed imaging system was utilized to capture bubble dynamics within the designated observation sections (Figure 1). The setup featured the same camera and a halogen backlight, ensuring uniform and high-contrast visualization of the bubbles. Imaging was performed at 0.64× optical magnification with a shutter speed of 3.33 $\mu$s, providing a spatial resolution of 2.9±0.1 $\mu$m over a 6×6 mm$^2$ field of view. Image sequences were recorded in lossless *.tif* format at 1 frame per second (fps) intervals, which both maximized image fidelity and minimized the likelihood of repeated bubble detection during subsequent post-processing.

### 3.2. *Bubble imaging – calibration, data processing and uncertainty analysis*

The imaging system was calibrated using a reference wire strand of known diameter (250±0.5 $\mu$m) positioned within a focal plane depth of 1.5±0.03mm. The resulting spatial resolution was ≈3 ±0.2 $\mu$m/pixel, and the estimated positional accuracy for clearly resolved bubble edges was ±6 $\mu$m. The smallest detected bubbles had a diameter of approximately 10 $\mu$m. For isolated, in-focus bubbles, this corresponds to a diameter uncertainty of ±10–12 $\mu$m, while in cases involving overlapping or partially defocused boundaries, manual verification and correction limited the maximum uncertainty to ±15 $\mu$m. Quantitative image analysis was performed using a custom MATLAB-based algorithm. Raw frames shown in Figure 3(a) were first preprocessed through background subtraction to suppress noise and enhance bubble edges. Bubble contours, as shown in Figure 3(b), were detected using the Canny edge detection method (Rong et al. 2014), and Watershed segmentation (Seal et al. 2015) was applied to resolve overlapping bubbles. Furthermore, the bubble size distribution for select images was cross-validated using the open-source, deep learning-based automated bubble detection algorithm developed by Kim & Park (2021). The projected two-dimensional area $A$ of each identified bubble was computed, and the equivalent bubble diameter (assuming spherical) was determined by assuming axisymmetry:

$$d = d_{\text{equiv}} = \left(\frac{4A}{\pi}\right)^{1/2}$$

Over 2000 bubbles were analyzed for each experimental condition to ensure statistical robustness. The Sauter mean diameter $d_{32}$ was computed as:

$$d_{32} = \frac{\sum_{i=1}^{N} n_i d_i^3}{\sum_{i=1}^{N} n_i d_i^2}, \qquad \text{and} \qquad \sum_{i=1}^{N} n_i = N,$$

where $n_i$ is the number of bubbles of diameter $d_i$, and $N$ is the total number of bubbles detected. To assess uncertainty in $d_{32}$, a sensitivity analysis was performed by varying segmentation thresholds and repeating the analysis on selected datasets. The relative uncertainty in $d_{32}$ was estimated to be ±4–6% (between repeated runs), primarily due to image processing variability and resolution limits. The assumption of bubble axisymmetry introduces a potential bias; however, based on the observed shapes (nearly spherical under moderate Weber number conditions), the associated error between the equivalent spherical and ellipsoidal areas remains within 5%, consistent with previous studies (Risso & Fabre 1998; Zenit & Magnaudet 2008; Razzaque et al. 2003b). To ensure reproducibility, each experimental condition was repeated three times on different days. The variation in results (e.g., $d_{32}$, bubble count) across these trials remained within the reported uncertainty bounds, confirming the consistency and robustness of the measurement and processing methodology.

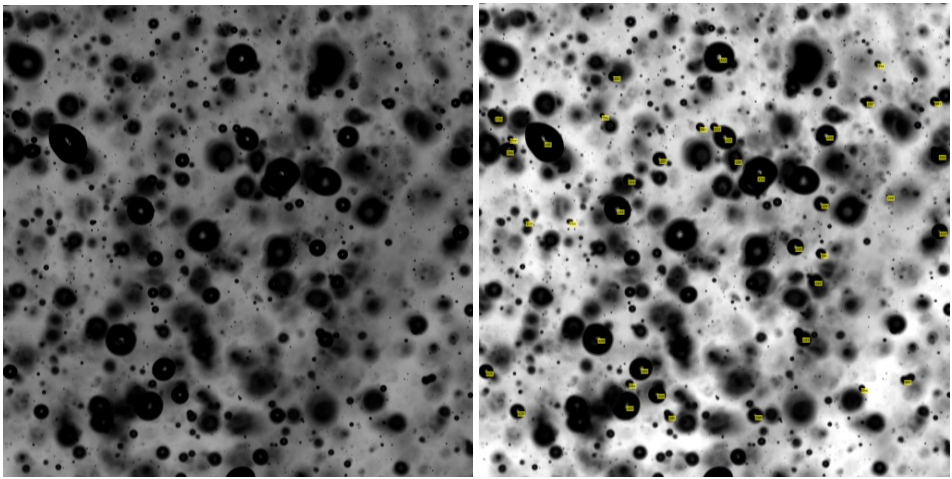

(a) Raw recorded image                    (b) Post-processed image

Figure 3. Image processing and bubble detection software (resized: 6×6 mm$^2$).

### 3.2.1. *Particle shadow velocimetry for turbulent field*

PSV provides a cost-effective, non-intrusive alternative to particle image velocimetry (PIV) for obtaining the flow velocity field (Estevadeordal & Goss 2005; Khodaparast et al. 2013; Hessenkemper & Ziegenhein 2018; Jassal et al. 2025). PSV was implemented to measure primarily the liquid-phase velocity in the vertical duct. A high-power Tungsten-Halogen light source (Thorlabs OSL2) back-illuminated the flow from behind (see Figure 1), casting sharp shadows of flow tracers onto the sensor of a high-speed Photron Nova R2 camera operated at 20000 fps. The depth of field of the lens used is ≈ 0.5 mm, which is comparable to the typical laser-sheet thickness in planar PIV applications. The flow was seeded with nearly neutrally buoyant hollow glass spheres (mean diameter ∼ 10 $\mu$m) which serve as PSV flow tracers. The response time of these tracers/particles is much smaller than the characteristic flow time scale, i.e. Stokes number ≪ 1, ensuring faithful tracking of the smallest turbulent motions. The calibrated field of view covered roughly 3 mm in the vertical direction at approximately 10 $\mu$m/pixel resolution. Note that the image resolution for PSV is lower than that for bubble imaging discussed previously. This was to capture sufficient particle and small bubble motions within the field of view. The PSV images (with both particle and bubble shadows) were processed, subtracting background and removing large bubble shadows, to convert them into formats suitable for cross-correlation based PIV processing (see Figure 4). Image pairs were then processed with the open-source software OpenPIV (Liberzon et al. 2020). We used 128×128 pixel interrogation windows with 75% overlap. Bubble shadows and any spurious large particles (high-Stokes tracers) were masked or excluded. The resulting instantaneous velocity fields were further checked and corrected for spurious data using the universal outlier detection method of Westerweel & Scarano (2005). The velocity statistics converged fully beyond 500 image-pairs and we conservatively used 1000 image-pairs. Compared to traditional laser-sheet PIV, PSV provides comparable accuracy for velocity and turbulence measurements in bubbly flows while avoiding complex optics and laser-light scattering issues (Jassal et al. 2025). The uncertainties in velocity and turbulence statistics were estimated using the methods of Wieneke (2015) and Sciacchitano & Wieneke (2016).

From the validated vector fields, we computed ensemble-averaged and fluctuation quantities, following the Reynolds decomposition of the velocity $u = \bar{u} + u'$. The mean velocity profile ($\bar{u}$) was obtained by ensemble-averaging each instantaneous vector component ($u$). The two-component turbulent kinetic energy ($tke$ or $k$) and turbulence intensity (TI) were calculated from the instantaneous fluctuations ($u'$), as shown below in Equation 3.1 to 3.3:

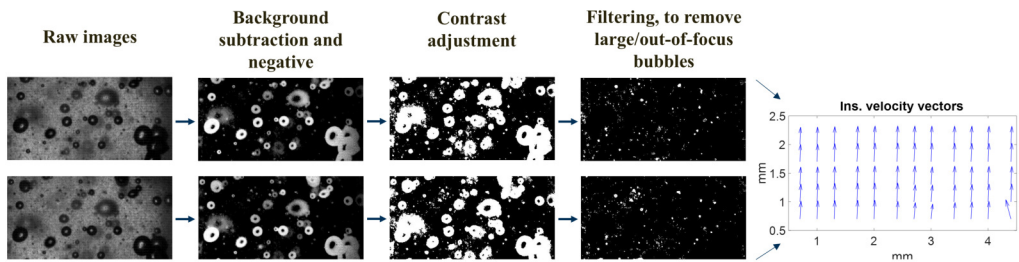

Figure 4. Different stages of image processing involved in particle shadow velocimetry.

$$\text{TKE} = tke = k = \frac{1}{2}\left(\overline{u'^2} + \overline{v'^2} + \overline{w'^2}\right) \tag{3.1}$$

$$\approx \frac{3}{4}\left(\overline{u'^2} + \overline{v'^2}\right), \quad \text{assuming } \overline{w'^2} \approx \left(\overline{u'^2} + \overline{v'^2}\right)/2 \tag{3.2}$$

$$\frac{\mathcal{U}}{V} = \text{TI} = \frac{\sqrt{\frac{2k}{3}}}{V} = \frac{1}{V}\sqrt{\frac{1}{3}\left(\overline{u'^2} + \overline{v'^2} + \overline{w'^2}\right)}, \quad \text{here } V \text{ is bulk velocity (6.1, 7.4, 8.4 m/s)} \tag{3.3}$$

The dissipation rate ($\varepsilon$) is estimated using

$$\varepsilon = 3\nu\left(\frac{5}{6}\left(\frac{\partial u'}{\partial x}\right)^2 + \frac{7}{12}\left(\frac{\partial v'}{\partial x}\right)^2 + \frac{7}{12}\left(\frac{\partial u'}{\partial y}\right)^2 + \frac{5}{6}\left(\frac{\partial v'}{\partial y}\right)^2 - \frac{\partial u'}{\partial x}\frac{\partial v'}{\partial y} - \frac{\partial v'}{\partial x}\frac{\partial u'}{\partial y}\right) \tag{3.4}$$

based on 2D velocity gradients under the local isotropy assumption (Xu & Chen 2013a; Verwey & Birouk 2022; Hinze 1975). Due to limited resolution of methods like PIV and PSV, the dissipation rates obtained from measured velocity fields and velocity gradients are considerably underestimated (Lavoie et al. 2007; Xu & Chen 2013b; Verwey & Birouk 2022). For the current work, we have used the structure function and spectra based empirical relationships of Xu & Chen (2013b) to correct our estimated dissipation rates.

## 4. Results and discussions

Section 4 provides a detailed characterization of turbulence and bubble dynamics in the developing duct flow. The analysis covers the axial evolution of turbulent kinetic energy ($k$) and dissipation rate ($\varepsilon$), followed by a comprehensive examination of bubble size distributions through the modified Gaussian function [$f(d)$], cumulative distribution [$\Phi(d)$], and their scaling behavior. Key statistical measures, including the Sauter mean diameter ($d_{32}$) and the extreme-to-mean size ratio ($d_{99.8}/d_{32}$), are quantified. Additionally, we study the influence of turbulence decay on the transition between coalescence- and breakup-dominated regimes, explore axial and radial variations in local void fraction ($\phi$), and establish scaling laws governing the evolution of bubble populations in high-intensity, decaying turbulence.

### 4.1. *Turbulent kinetic energy and dissipation rate*

The evolution of the turbulent and velocity fields along the duct axis is critical for understanding bubble dynamics in developing flows. This spatial variation in turbulence directly governs the interplay between bubble breakup and coalescence, making its quantification essential for predicting bubble size evolution and optimizing industrial multiphase flow systems (Li & Liao 2024). Figures 5 and 6 present the axial evolution of centerline turbulent kinetic energy ($k$) and turbulent dissipation rate ($\varepsilon$) at different bulk velocities ($V$).

In all cases, the turbulence intensity imparted by the pump rapidly decreases as the flow develops; the centerline $k$ (normalized by $V^2$) and $\varepsilon$ both decrease by an order of magnitude (over $\sim 90\%$ reduction) from the inlet to the farthest measurement station. This steep initial decay reflects the absence of sustaining turbulence production in the core flow immediately after the pump, so the injected turbulent eddies dissipate energy quickly. Beyond $\approx 22$ hydraulic diameters downstream, the rate of decline becomes much gentler (almost leveling off for the lowest bulk velocity case), indicating the approach toward a quasi-equilibrium turbulence state. At the lowest bulk velocity $V = 6.1$ m/s, for instance, $\varepsilon$ drops very rapidly in the first few hydraulic diameters and then flattens by $\mathcal{L} \approx 30$ (suggesting that the core turbulence has largely equilibrated by that distance). In contrast, at the highest velocity ($V = 8.4$ m/s), centerline dissipation is still gradually decreasing even at the farthest location ($\mathcal{L} \approx 40$), implying a longer development length before fully developed conditions are reached. These observed trends – an initial exponential-like decay of turbulence followed by a slower power law-like tail – are consistent with classical decaying turbulence behavior downstream of intense turbulence sources, and they align with prior findings that high core turbulence decays rapidly in developing pipe flows until wall-driven production begins to dominate (Doherty et al. 2007; Wilson & Smith 2007).

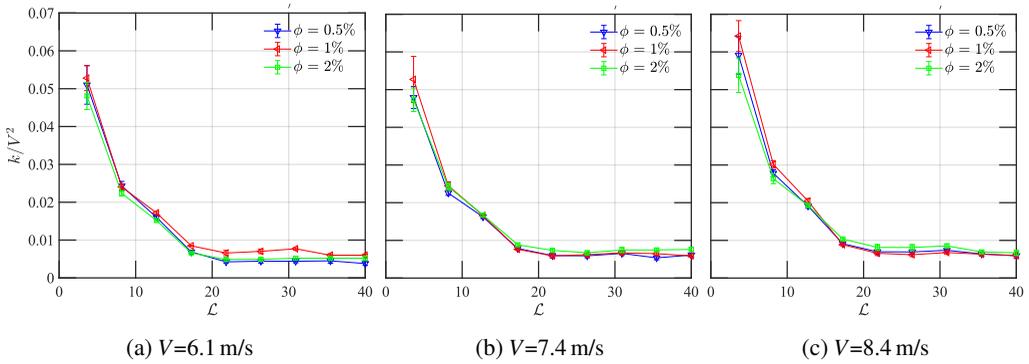

Figure 5. Axial variation of *tke* ($k$) along the duct centerline for three bulk velocities $V(Q)$ and three void fractions $\phi$.

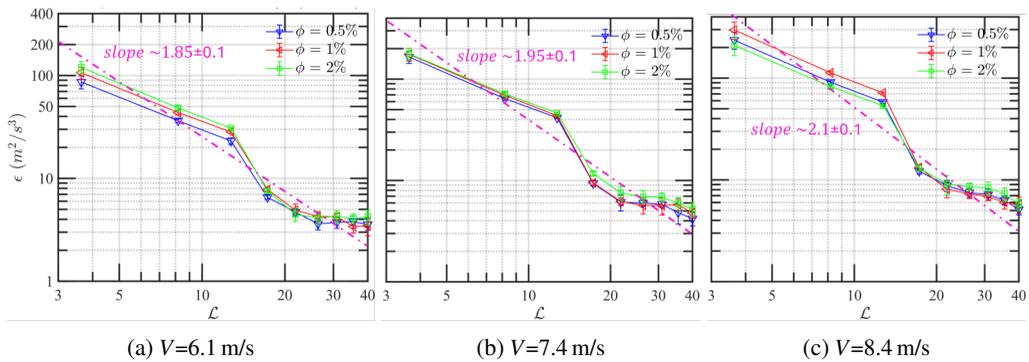

Figure 6. Axial variation of $\varepsilon$ along the duct centerline, for three bulk velocities $V(Q)$ and three void fractions $\phi$. Note the use of log-log scale.

Increasing $V$ significantly elevates both the centerline $k$ and $\varepsilon$ across the board. Flows at higher Re inject more energy into turbulence, yielding larger initial $k$ and accordingly, higher $\varepsilon$ values at the inlet and downstream. For instance, at $\mathcal{L} \approx 5$ the dissipation in the $V = 8.4$ m/s case is noticeably greater than in the $V = 6.1$ m/s case for the same air void fraction $\phi$, and this gap persists downstream, although all cases eventually decay to significantly lower levels. Such a trend has been reported in other systems – e.g., Shawkat et al. (2008) observed that the turbulence dissipation in horizontal bubbly flows rises with increasing $V$. Notably, the higher Re cases not only start with larger $\varepsilon$ near the inlet, but also retain a higher dissipation level further downstream (since a more energetic flow takes longer to dissipate). The fact that the $V = 8.4$ m/s data in Figure 5(c) still show a slight downward slope at $\mathcal{L} = 40$, whereas the $V = 6.1$ m/s data in Figure 5(a) have nearly plateaued by that distance, suggests that the development length to reach a fully developed turbulence profile grows with Re. This is consistent with single-phase pipe flow literature – higher inertia flows require longer distances for turbulence redistribution and stabilization, underscoring the significant influence of Re on turbulence evolution in our experiments (Doherty et al. 2007; Wilson & Smith 2007).

For the relatively low void fractions ($\phi = 0.5\%$ - $2\%$) in the current study, the influence of $\phi$ on turbulence statistics is quite subtle. Any changes in the measured turbulent kinetic energy $k$ and dissipation rate $\varepsilon$ across this range remain within experimental uncertainty, in part because the PSV system lacks the fine-scale resolution to capture such small variations. Indeed, as shown in Figures 5 and 6, the $k$ and $\varepsilon$ profiles at these low void fractions $\phi$ nearly overlap. This suggests that a threshold void fraction is required before bubbles appreciably augment the turbulence – a notion supported by prior studies. Lance & Bataille (1991) identified a regime at low void fraction where bubble–turbulence interactions are negligible, and Serizawa et al. (1975) likewise observed minimal turbulence modification at void fractions on the order of 1%. Similarly, Shawkat et al. (2008) reported that at void levels of 1%, any bubble-induced changes in turbulence are minor and often within the measurement scatter. Under the dilute conditions of the present study, therefore, the net impact of $\phi$ on $k$ and $\varepsilon$ is modest, and subtle bubble-induced enhancements cannot be reliably distinguished from experimental noise.

At the duct inlet, where the flow emerges directly from the pump, the turbulent kinetic energy imparted by the pump is initially distributed relatively uniformly across the pipe cross-section, resulting in elevated turbulent dissipation rates both at the center and near the wall (refer appendix A and B). This high core dissipation arises from the injection of large-scale, isotropic turbulent eddies that permeate the bulk flow, a phenomenon frequently observed in turbulent flows downstream of energetic mechanical sources (Pope 2000; Durst et al. 1995). In this region, turbulence production is not yet dominated by wall-shear effects, and the dissipation profile lacks the classical near-wall peak that is characteristic of fully developed pipe flows (Eggels et al. 1994). As the flow proceeds downstream, the imposed turbulence in the pipe core decays rapidly due to the absence of a sustaining production mechanism away from the wall (Pope 2000). In contrast, near-wall turbulence decays slower and is withheld by the strong mean velocity gradient at the solid boundary, which becomes the primary source of turbulent kinetic energy production and subsequent dissipation (Durst et al. 1995). Consequently, the radial dissipation profile evolves to exhibit a pronounced peak near the wall and a significant reduction at the centerline, reflecting the gradual transition from pump-driven, spatially uniform turbulence to a fully developed, wall-dominated turbulent regime (Eggels et al. 1994). This observed evolution aligns with established theoretical and experimental findings on the development of turbulence in wall-bounded shear flows and underscores the interplay between initial isotropic turbulence and the emergence of wall-driven turbulent structures.

Figure 6 shows that the turbulent dissipation decays approximately as a power law ($\varepsilon \propto \mathcal{L}^{-m}$) along the duct, appearing linear on a log–log plot with a best-fit slope $m$ narrowly confined to $m \simeq 1.85 - 2.10$. The spread is small enough that the corresponding error bars largely overlap. This behavior holds over most axial stations, except near the end where the growing wall boundary layer sustains turbulence and the decay becomes noticeably slower. As we see weak radial variations in our measurements, this scaling can be compared with theory by assuming homogeneous isotropic turbulence (HIT). In decaying HIT, the $\varepsilon$ scaling follows a power law whose exponent is dependent on the large-scale invariant: where $k l^p = $ constant with $p \in \{3, 5\}$ (Saffman 1967; Batchelor & Proudman 1956). This implies

$$k \, l^p = \text{constant} \quad \Rightarrow \quad l \propto k^{-1/p}.$$

where $l$ is integral length scale. For HIT,

$$\varepsilon \sim C_\varepsilon \frac{k^{3/2}}{l} \; \Rightarrow \; \varepsilon \propto k^{3/2 + 1/p}, \qquad \frac{dk}{dt} = -\varepsilon \; \Rightarrow \; \frac{dk}{dt} \propto -k^m,$$

where t is time, and $m = \frac{3}{2} + \frac{1}{p}$. Integration gives

$$k(t) \propto t^{-\frac{1}{m-1}} = t^{-\frac{2p}{p+2}}.$$

$$l(t) \propto k^{-1/p} \propto t^{\frac{2}{p+2}}, \qquad \varepsilon(t) \propto \frac{k^{3/2}}{L} \propto t^{-\frac{3p+2}{p+2}}.$$

Simplifying for the case $t \to \mathcal{L}/V$ gives,

$$\varepsilon(t) \propto t^{-m} \propto \mathcal{L}^{-m}, \qquad m = \frac{3p+2}{p+2}. \tag{4.1}$$

Thus, for a Saffman (1967) $\kappa^2$ spectrum ($p = 3$; permanence of big eddies for isotropic energy spectrum, $E(\kappa) \sim \kappa^2$ as $\kappa \to 0$; here $\kappa$ is wave number), one gets $\varepsilon \sim \mathcal{L}^{-2.2}$, while for a Loitsyanskii (1939) and Batchelor & Proudman (1956) $\kappa^4$ spectrum, ($p = 5$) one gets $\varepsilon \sim \mathcal{L}^{-2.43}$. These theoretical scalings are broadly consistent and closely matches with experimental measurements (refer Figure 6). A small but systematic discrepancy remains: the experimentally observed decay is slightly slower than the ideal HIT prediction, which is primarily due to wall-bounded effects in the duct that sustain dissipation and retard the streamwise decay, more pronounced near the end of the duct.

### 4.2. *Bubble size distribution and flow regime evolution*

The modified Gaussian function $[f(d)]$ is widely used to compare and characterize the bubble-size population: it locates where most bubbles reside (via the mode or geometric mean), bounds the largest sizes (e.g., $d_{max}$), and quantifies dispersion about the mean (standard deviation). Controlling the mean/mode, $\sigma$, and $d_{max}$ of the bubble-size distribution in the flow is critical for (i) chemical bubble/slurry reactors (e.g., hydrogenation, Fischer–Tropsch) (Kulkarni 2007), (ii) bioreactors and wastewater aeration (Garcia-Ochoa & Gomez 2009; Kantarci et al. 2005), (iii) mineral flotation/DAF (Wang et al. 2020), and (iv) $CO_2$ absorption/scrubbing (Chen et al. 2023), etc. The $f(d)$ is represented by Equation 4.2 and 4.3:

$$f(d) = \frac{\partial \Phi}{\partial \ln d} = \frac{1}{\sqrt{2\pi} \ln \sigma_g} \exp\left[ -\frac{1}{2} \left( \frac{\ln(d/d_g)}{\ln \sigma_g} \right)^2 \right] \tag{4.2}$$

$$\text{where} \quad d_g = \left( \prod_{i=1}^{K} (d_i)^{n_i} \right)^{\frac{1}{K}} \quad \text{and} \quad \ln \sigma_g = \sqrt{\frac{\sum_{i=1}^{K} n_i (\ln d_i - \ln d_g)^2}{N}} \tag{4.3}$$

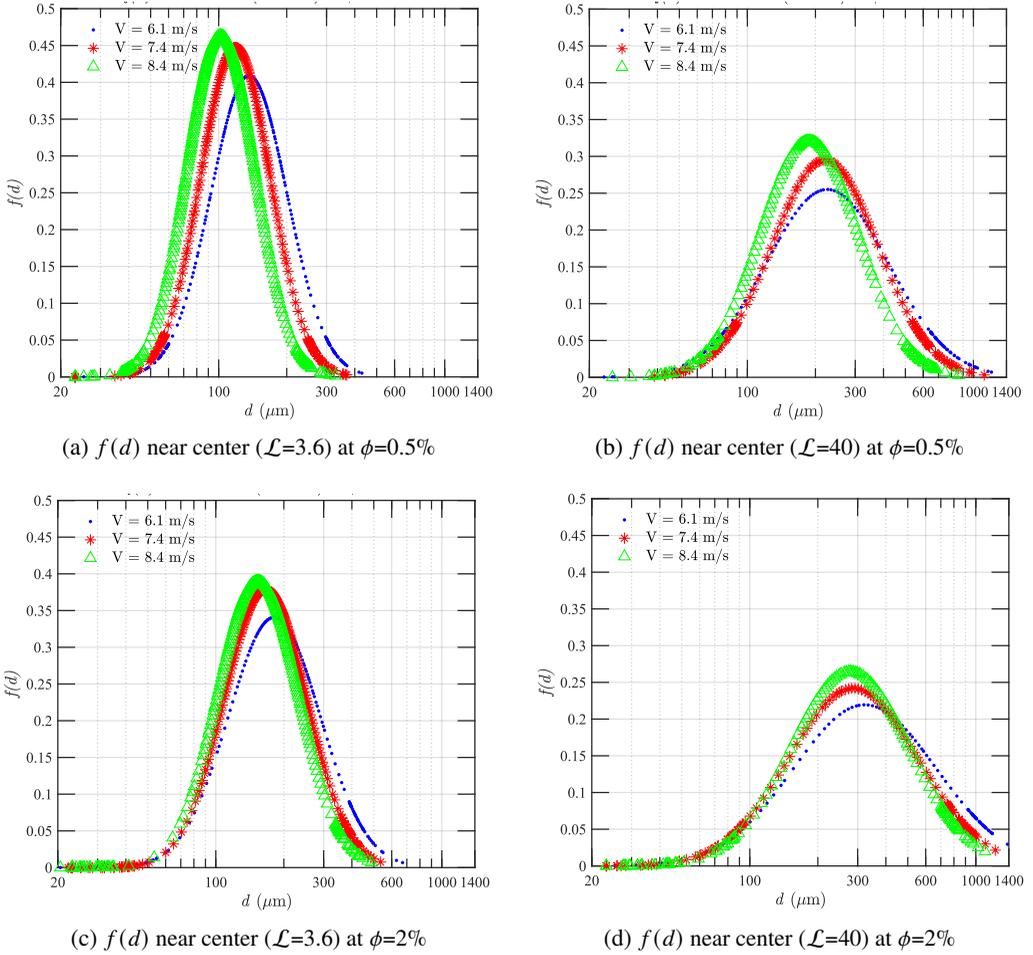

(a) $f(d)$ near center ($\mathcal{L}$=3.6) at $\phi$=0.5%

(b) $f(d)$ near center ($\mathcal{L}$=40) at $\phi$=0.5%

(c) $f(d)$ near center ($\mathcal{L}$=3.6) at $\phi$=2%

(d) $f(d)$ near center ($\mathcal{L}$=40) at $\phi$=2%

Figure 7. Bubble size distribution with modified Gaussian function $f(d)$.

The parameters here are the geometric mean diameter ($d_g$) and geometric standard deviation ($\sigma_g$) (Razzaque et al. 2003b). Here, $\Phi$ is the cumulative fraction of bubbles that have diameters smaller than a given diameter, $d$. The Figure 7(a, b, c & d) shows that increasing velocity ($V$) (or Re) shifts the normalized bubble-size distribution toward smaller mean and standard deviation of bubble size distribution. At the highest $V$ (8.4 m/s), $f(d)$ exhibits a tall, sharp peak at a relatively small diameter, with a greatly diminished tail of large bubbles. In contrast, at the lowest $V$ (6.1 m/s) the distribution is broader and more right-sided, with a noticeable tail consisting of larger bubbles. Physically, the more energetic turbulence at higher $\varepsilon$ breaks bubbles into smaller sizes and rapidly fragment any large bubbles, yielding a narrower size spectrum (Prince & Blanch 1990a). At lower $\varepsilon$ the turbulent stresses are weaker, so some larger bubbles can survive without breaking up, producing a wider distribution. These trends are consistent with the idea that turbulent breakup dominates at high dissipation rates while permitting large-bubble persistence at lower turbulence levels (Prince & Blanch 1990a; Lance & Bataille 1991). In fact, classic coalescence-breakup models of Prince & Blanch (1990a) treat turbulence as the driver of bubble collisions, but sufficient contact time is required for coalescence. Extremely intense turbulence shortens bubble contact times, suppressing coalescence efficiency, so breakup prevails and limits the upper bubble size. This is in agreement with the observed disappearance of the large-diameter tail at the highest velocity.

Moving downstream in the duct (from $\mathcal{L} \simeq 3.6$ to $40$ after injection) shown in Figure 7($a \rightarrow b$, $c \rightarrow d$), the bubble size distribution progressively broadens and develops a less pronounced peak. Near the injection point ($\mathcal{L} \simeq 3.6$ downstream), $f(d)$ is relatively narrow with a small tail of both very small and a few large bubbles – a signature of the initial breakup of the air flow by the pump's turbulence. With an increase in $\mathcal{L}$ (shown for $\mathcal{L} = 40$ in Figure 7), the distribution becomes broader around a modal diameter, and the extreme (specifically very large bubbles) are more prevalent. This broadening of the distribution reflects the combined effects of decaying turbulence and ongoing bubble–bubble interactions. As the turbulence energy dissipates downstream, fewer new micro-bubbles are generated by breakup, and the small bubbles present begin to coalesce into mid-sized ones. The net result is an evolving equilibrium toward coalescence: the overall bubble count decreases with distance, indicating that many small bubbles are merging into larger ones. Notably, the large-bubble tail of the distribution grows significantly by $\mathcal{L} \sim 40$, suggesting that large bubbles are formed via coalescence in the still-turbulent (though weakening) flow. Thus, over distance the population shifts toward higher bubble diameter. This trend of an initially narrow distribution moving toward a broader and smaller peaked distribution downstream has not been observed in other developing bubbly flows. However, prior studies show a coalescence-dominated regime—with a bubble-size distribution biased to larger bubbles—in fully developed horizontal pipe flow (Razzaque et al. 2003b) and in decaying turbulence (Serizawa et al. 1975). Our findings here similarly indicate that in axial direction, the turbulence-decaying, coalescence-dominated regime yields a lower peak and broader $f(d)$ distribution.

Higher $\phi$ leads to a broader and flatter bubble size distribution as shown in Figure 7($a \rightarrow c$, $b \rightarrow d$), with an enhanced probability of larger bubbles. In these higher void fraction cases, the distribution's large-diameter tail becomes more pronounced, meaning a greater presence of big bubbles compared to lower void conditions. The physics driving this trend is the increased frequency of bubble–bubble collisions at higher $\phi$. Even though the flow is turbulent, a higher collision rate allows many small bubbles to merge into larger ones and decaying turbulence provides stability. Although rapidly decaying, developing turbulent multiphase flows are scarcely examined, our observations align with results for fully developed, coalescence-prone turbulence: increasing the void fraction $\phi$ raises the mean bubble size and broadens the distribution (Lehr et al. 2002a). At high $\phi$, collision probabilities—and hence collision/coalescence frequencies—increase (Prince & Blanch 1990b; Razzaque et al. 2003b), directly shaping the size–distribution function $f(d)$. By contrast, in strongly breakup-dominated regimes, bubble size can remain nearly independent of $\phi$ up to a threshold. In our measurements, coalescence plays a measurable role: as $\phi$ increases, enhanced bubble–bubble interactions shift $f(d)$ toward larger diameters and yield a broader, right-sided distribution, consistent with Lance & Bataille (1991).

### 4.3. *Scaling of bubble size distribution*

While mean diameters and distribution widths characterize bulk bubble evolution, the normalized probability density function (*pdf*) of sizes captures the complete statistical structure of the population. The *pdf* provides physical characterization of the distribution: the presence and slope of power law tails and their axial evolution - identify the dominant mechanism (inertial/capillary breakup vs. coalescence) and delineate regime transitions (Deane & Stokes 2002a; Chan et al. 2018a). Figure 8 compares the bubble size *pdf* measured near the duct centerline at two axial locations ($\mathcal{L} = 3.6$ & $40$) for two void

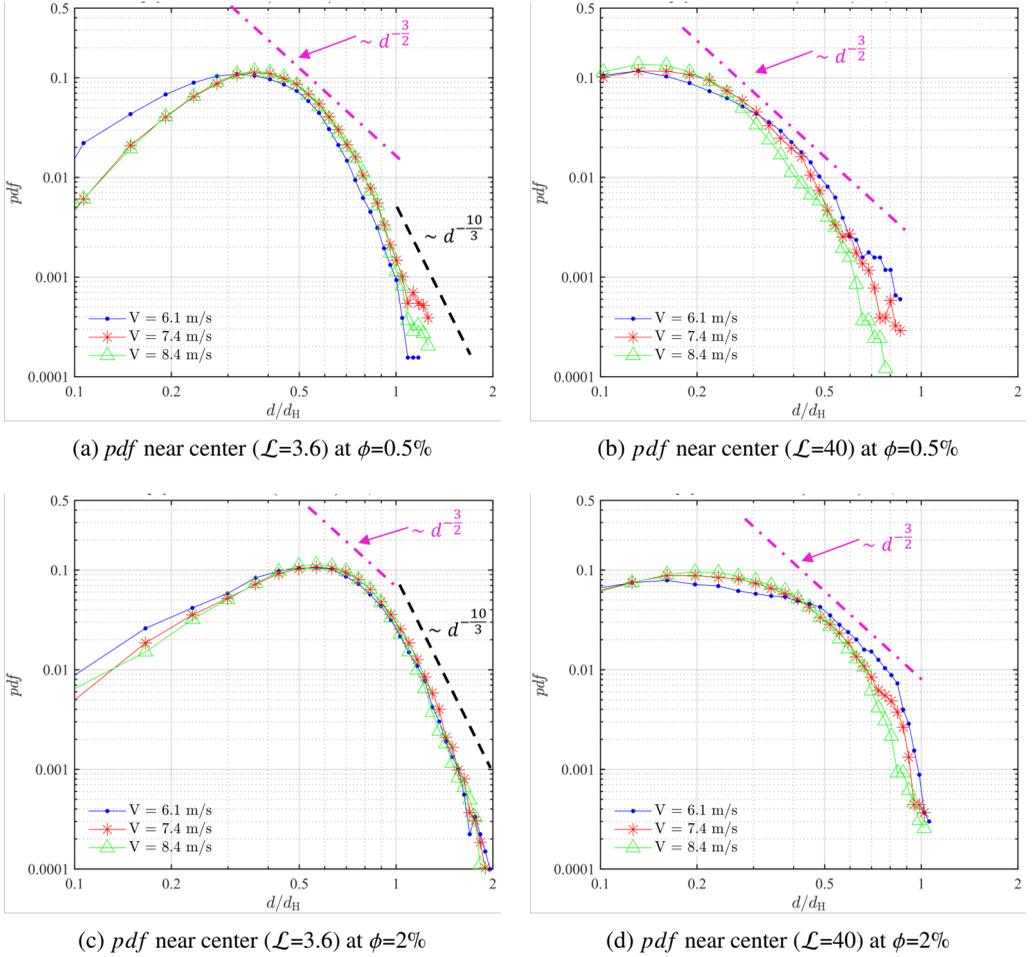

(a) *pdf* near center ($\mathcal{L}$=3.6) at $\phi$=0.5%

(b) *pdf* near center ($\mathcal{L}$=40) at $\phi$=0.5%

(c) *pdf* near center ($\mathcal{L}$=3.6) at $\phi$=2%

(d) *pdf* near center ($\mathcal{L}$=40) at $\phi$=2%

Figure 8. Scaling of probability density function (*pdf*). Bubble size $d$ is normalized with Hinze scale $d_{\mathrm{H}}$. Magenta power law scaling: $d^{-3/2}$; Black power law scaling: $d^{-10/3}$.

fraction ($\phi$ = 0.5% & 2%) and varying bulk velocities (or $Q$). Near the inlet ($\mathcal{L}$ = 3.6), both void fraction cases (Figures 8(a) and 8(c)) exhibit a similar dual-slope pdf behavior on log-log plots. Bubble diameters in *sub-Hinze* follow an approximate $d^{-3/2}$ power law, indicating a no active breakup regime (mostly coalescence) (Crialesi-Esposito et al. 2023). In this region ($\mathcal{L}$ = 3.6, for all $\phi$ and $V$) virtually all of the bubble population lies under the Hinze diameter ($d/d_{\mathrm{H}} < 1$), so turbulent eddies are too weak to overcome surface tension and break the bubbles. Consistently, breakup is relatively small at $\mathcal{L}$ = 3.6 but bubble growth is governed by both coalescence and breakup. The observed $-3/2$ scaling is consistent with prior findings by Deane & Stokes (2002) in flows lacking active fragmentation, where this exponent is classically associated with *sub-Hinze* bubble generation via capillary mechanisms observed in breaking waves or turbulent-induced pinch-off events (Deane & Stokes 2002b). For example, oceanic wave-breaking experiments showed that *sub-Hinze* bubbles obey a $d^{-3/2}$ size spectrum, similar to our inlet measurements (Deane & Stokes 2002b; Farsoiya et al. 2023; Roccon et al. 2023; Jain & Elnahhas 2025). This finding also accords with the notion that as long as bubbles remain smaller than the critical Hinze size for breakup, successive collisions will merge them into larger bubbles and broaden the distribution without any competing fragmenting mechanism (Wang et al. 2016). Our data indeed provide direct evidence of such a coalescence-driven cascade: an initially narrow bubble size distribution at the duct inlet rapidly evolves into a broader $d^{-3/2}$ distribution before any significant breakup occurs. Notably, this near inlet $d^{-3/2}$ pdf holds for all tested flow conditions – all three liquid velocities and both void fractions collapse to the same slope at $\mathcal{L} = 3.6$ – underscoring that the early developing flow is a robust coalescence-dominated regime largely independent of $\phi$ or turbulence intensity (Prince & Blanch 1990b; Farsoiya et al. 2023). Although most of the bubbles are *sub-Hinze* scale, there are considerable amounts of bubbles with *super-Hinze* scale ($d > d_H$) at $\mathcal{L} = 3.6$ due to still high $\varepsilon$. These *super-Hinze* scale bubbles are susceptible to turbulent breakup, causing the pdf's upper tail to steepen toward a $\sim d^{-10/3}$ slope. Figure 8 shows that once bubbles exceed the local $d_H$, they indeed begin to fragment and the distribution tail assumes the characteristic $\sim d^{-10/3}$ form associated with inertial breakup. A power law fit to the largest bubbles at $\mathcal{L} = 3.6$ yields an exponent close to $-3.8$, closely matching the expected $-10/3$ scaling for inertial breakup. This $d^{-10/3}$ scaling for fragmenting bubbles is consistent with classical fragmentation cascade models (Garrett & Li 2000) and has been observed in experiments and simulations of turbulent dispersions (Soligo et al. 2019). It is worth noting that due to the relatively low $\phi$, the population of these large bubbles is sparse – the tail can appear even steeper than $-10/3$. This validates the trend also noted by Farsoiya et al. (2023) in recent direct numerical simulations for low $\phi$.

Far downstream ($\mathcal{L} = 40$), the bubble size distribution has shifted and developed a pronounced single slope scaling. All bubbles remain *sub-Hinze* and continue to follow the coalescence dominated $d^{-3/2}$ trend. Unlike an equilibrium regime where breakup balances coalescence effects, here coalescence still dominates the majority of the duct region, while limited breakup occurs only near the inlet ($\mathcal{L} < 5-8$, depending on $V$) where $\varepsilon$ is very high (more details in Section 4.5). In fact, beyond the $\mathcal{L} > 5-8$ downstream of the inlet, turbulent energy has decayed so substantially that *super-Hinze* scale bubble breakup becomes non-existent – the flow effectively transitions into a nearly purely coalescence-driven regime (more details in Section 4.5). The disappearance of the *super-Hinze* tail with increasing $\mathcal{L}$ (from $\mathcal{L} = 3.6$) indicates that while some bubbles briefly exceed the Hinze scale, they are rapidly broken up, preventing a persistent *super-Hinze* population. This behavior is fully consistent with the concept of the Hinze scale acting as a cutoff: a critical diameter $d_H$ exists at which turbulent dynamic pressure balances capillary pressure ($\rho \mathcal{V}^2 \sim \gamma/d_H$), corresponding to a Weber number of order unity ($We \sim O(1)$) for breakup onset (Chan et al. 2018b; Hinze 1955b). This equilibrium size concept has long been utilized in population balance models to predict stable bubble size distributions in steady-state bubbly flows (Lehr et al. 2002a). In our decaying flow, however, the continuous reduction in turbulent intensity with increasing $\mathcal{L}$ ensures that bubbles rarely exceed the Hinze scale. Bubbles that do exceed the Hinze scale are not actively fragmented. Consequently, beyond a few hydraulic diameters from the inlet, bubble coalescence mostly proceeds unopposed by breakup.

The influences of $\phi$ and $V$ on the pdf scaling trends are also evident in Figure 8. Near inlet ($\mathcal{L} \leq 8.2$), cases with higher void fraction ($\phi = 2\%$ vs. $0.5\%$) exhibit a broader distribution due to accelerated coalescence-driven bubble growth, along with a more pronounced tail of large bubbles following the $d^{-10/3}$ scaling (comparing Figures 8(c) and 8(a)). Although at $\mathcal{L} = 40$, the $2\%$ case exhibits predominantly a single-slope regime, it still displays a longer large-bubble tail compared to the $0.5\%$ case - again attributable to the enhanced coalescence at higher $\phi$ (comparing Figures 8(d) and 8(b)). This trend aligns with theoretical expectations: a higher concentration of bubbles increases the frequency of collisions and consequently the coalescence rate, allowing bubbles to reach the critical Hinze scale more rapidly (Wang et al. 2016). In practical terms, increasing the void fraction enhances the extent of the distribution and leads to earlier formation of large, unstable bubbles susceptible to breakup within a few hydraulic diameters from the inlet - manifesting as an extended $d^{-10/3}$ fragmentation tail. Downstream of this region, even as the turbulence continues to decay, the higher $\phi$ case maintains a larger population of bubbles and a longer $d^{-3/2}$ tail relative to the lower $\phi$ case.

Bulk liquid velocity exerts a subtler yet important influence on the pdf scaling. Remarkably, for a given void fraction $\phi$, the normalized distributions, plotted as pdf versus $d/d_\mathrm{H}$ - collapse onto one another, showing strong overlap across different axial locations $\mathcal{L}$ and velocities $V$. This collapse suggests a self-similar evolution of the bubble size distribution when scaled by the local Hinze diameter. The behavior can be understood by examining the role of $\varepsilon$: higher liquid velocities inject more turbulent energy, leading to greater dissipation rates ($\varepsilon$) and consequently smaller Hinze scales ($d_\mathrm{H}$). In these high-$V$ cases, strong turbulence suppresses the coalescence-driven growth (due to reduced coalescence timescales) of large bubbles by fragmenting them early, effectively "nipping in the bud" any transition toward a broader distribution (Prince & Blanch 1990a). As a result, the bubble population in high-$V$ flows remains narrowly distributed and biased toward smaller diameters, with minimal spread around the mean. Conversely, lower-velocity flows (e.g., $V = 6.1$ m/s) exhibit weaker turbulence (smaller $\varepsilon$) and thus larger Hinze diameters (Yang et al. 2025). Under such conditions, bubbles can grow to larger sizes before turbulent stresses are sufficient to induce breakup, resulting in a broader size distribution and a more prominent large-bubble population. Although the low-$V$ flows still experience decaying turbulence and correspondingly smaller $d_\mathrm{H}$ farther downstream, the diminished fragmentation enables broader coalescence-driven growth. *These results confirm that the bubble size distribution in decaying turbulent flows scales robustly with the local Hinze diameter, and the normalized pdfs exhibit a self-similar trend across a range of flow conditions.*

### 4.4. *Evolution of the cumulative bubble size distribution ($\Phi$) in developing turbulent flows*

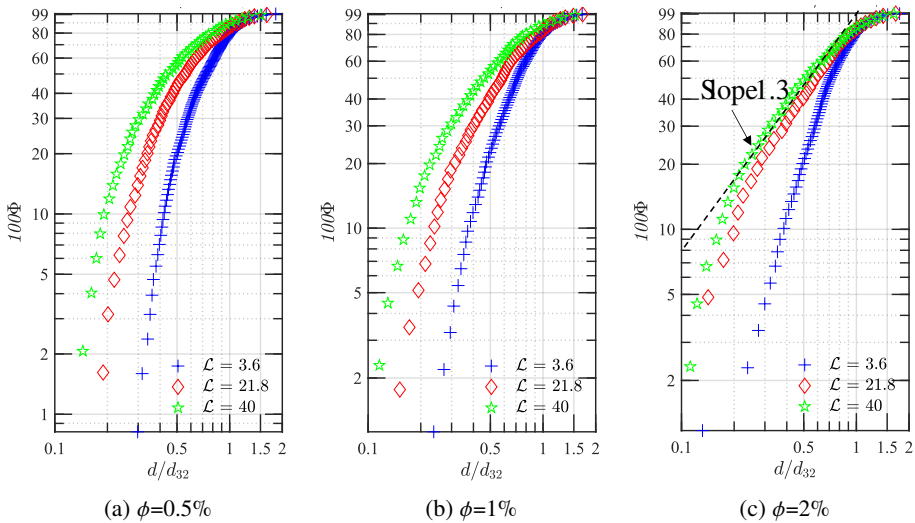

(a) $\phi$=0.5%  (b) $\phi$=1%  (c) $\phi$=2%

Figure 9. Cumulative bubble size distribution ($\Phi$) with bubble size ($d/d_{32}$) at $V = 6.1$ m/s.

Characterizing bubble size distribution in gas-liquid turbulent flows is essential for predicting and controlling key process parameters, such as mass transfer rates and interfacial areas in multiphase systems (Prince & Blanch 1990b; Lehr et al. 2002a). The cumulative bubble size distribution function, represented as $\Phi$, plotted against the normalized bubble diameter $d/d_{32}$, provides a robust means to describe bubble population dynamics, particularly elucidating transitions from non-equilibrium toward fully developed equilibrium states (Razzaque et al. 2003b). While previous studies have primarily focused on fully developed turbulent bubbly flows, comprehensive experimental characterizations in highly turbulent, developing regimes remain limited - an area explicitly addressed in this study.

Figure 9 presents the cumulative bubble size distributions measured at a bulk velocity of 6.1 m/s across varying initial void fractions (0.5%, 1%, and 2%). Measurements are presented at distinct axial positions ($\mathcal{L} = 3.6$, 21.8, 40) to illustrate the axial evolution in a coalescence-dominated flow regime. Near the duct inlet ($\mathcal{L} = 3.6$), the cumulative distribution $\Phi$ exhibits pronounced nonlinearity, with steep slopes for small bubbles and progressively declining slopes for larger bubbles. This reflects a population dominated initially by bubbles significantly smaller than the equilibrium bubble size under the given flow conditions. As the flow travels downstream, the cumulative distributions progressively flatten, indicating the preferential growth of smaller bubbles into larger bubbles through enhanced coalescence. This transition is notably accelerated at higher void fractions, revealing a clear correlation between increased $\phi$ and coalescence rate.

At the highest void fraction tested (2%), the cumulative distribution achieves near-linear form (on log-log) at downstream locations ($\mathcal{L} = 21.8$ and $\mathcal{L} = 40$), signifying a bubble size distribution nearing equilibrium - a condition previously associated exclusively with fully developed turbulent bubbly flows (Razzaque et al. 2003b). The pronounced nonlinearity observed near the duct inlet results primarily from bubble injection processes combined with intense initial turbulence. Initially small bubbles, rapidly formed due to vigorous turbulent breakup at the pump discharge, dominate the inlet distribution, as demonstrated by the steep slopes at low bubble diameters (Prince & Blanch 1990b; Razzaque 2005). Such conditions deviate from classical equilibrium distributions, which typically show uniform slopes when plotted in cumulative form (Razzaque et al. 2003b; Lehr et al. 2002a). Downstream progression clearly demonstrates the interplay of turbulence decay and bubble coalescence mechanisms. As turbulence intensity declines, small bubbles experience prolonged interactions leading to frequent coalescence events, reflected by decreasing slopes in cumulative distributions. Literature consistently supports the concept that turbulence reduction downstream from high-energy inlet conditions leads to increased bubble growth via coalescence, eventually stabilizing toward equilibrium distributions (Chieco & Durian 2023; Chan et al. 2018a).

Notably, our experiments identify a clear relationship between void fraction and the rate at which distributions approach equilibrium. The accelerated flattening of cumulative distributions at higher void fractions corroborates observations by Liao & Lucas (2010b) and Lehr et al. (2002a), who independently demonstrated that increasing bubble concentrations dramatically enhance collision frequencies and coalescence probabilities. At the highest tested void fraction (2%) in our study, this accelerated bubble coalescence rate rapidly drives the distribution toward near-linearity by mid-duct ($\mathcal{L} = 21.8$), indicating an almost fully developed equilibrium profile. Conversely, lower void fractions require significantly longer distances to achieve similar equilibrium, highlighting the critical role of bubble number density in coalescence dynamics.

### 4.5. *Evolution of $d_{32}$, $d_{\mathrm{H}}$ and $d_{\max}$ with $\mathcal{L}$: coalescence vs. turbulence-limited growth*

The axial evolution of the bubble-size distribution, combining $d_{32}$ with high quantiles such as $d_{99.8}$ (99.8th percentile bubble diameter) and referencing the Hinze scale $d_{\mathrm{H}}$, captures both bulk and tail dynamics as turbulence decays. Identifying shifts between breakup-limited and coalescence-dominated regimes is critical, since the former suppresses large bubbles while the latter promotes tail broadening and enhanced interfacial renewal. Such regime distinctions directly impact performance in electrochemical reactors by stabilizing mass-transfer layers and current density (Taqieddin et al. 2017; Angulo et al. 2020), in gas-lift risers and multiphase pipelines by delaying slug initiation and regulating pressure drop (Diaz et al. 2024; De Temmerman et al. 2015), and in membrane air-scouring by maximizing near-wall renewal at fixed aeration power (Alameedy et al. 2025; Fabre & Liné 1992). Regime-resolved distribution tracking therefore, provides stringent calibration targets for population-balance models and a mechanistic basis for reliable scale-up.

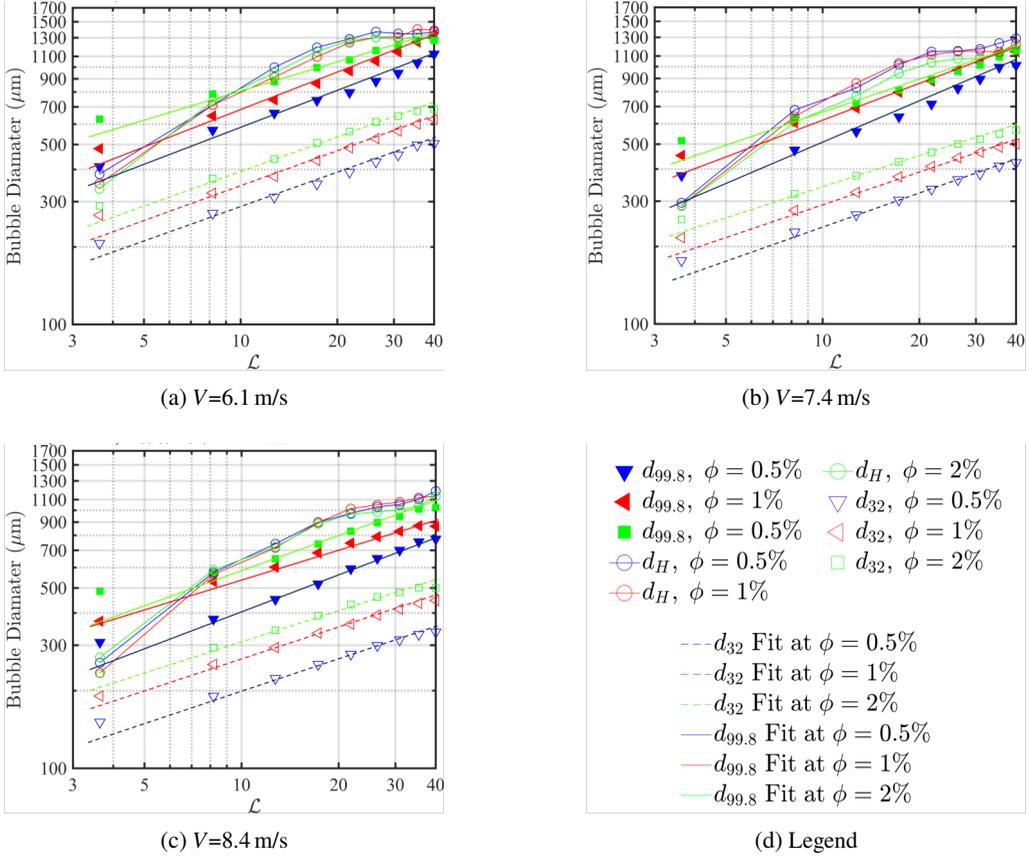

(a) $V$=6.1 m/s

(b) $V$=7.4 m/s

(c) $V$=8.4 m/s

(d) Legend

Figure 10. Axial variation of $d_{99.8}$, $d_{32}$, and $d_H$ in developing turbulent duct flow. The reported fit parameters $\beta_{\exp}$ (power law coefficient) and $R^2$ (coefficient of determination) are obtained from data for $\mathcal{L} > 3.6$, excluding the first data point, and are reported in Table 1.

Figure 10 presents the axial variation of bubble size statistics on a log-log scale, namely $d_{32}$, $d_{99.8}$, and the Hinze scale $d_H$, for each void fraction $\phi$ and bulk velocity $V$. The Hinze scale is evaluated locally at each axial station from Equation in section 4.4 (Hinze 1955a):

$$d_{\mathrm{H}} = We_{\mathrm{crit}}^{3/5} \left( \frac{\gamma}{\rho_l} \right)^{3/5} \varepsilon^{-2/5} \tag{4.4}$$

where $\gamma$ is the surface tension and $\rho_l$ is the liquid phase density. The critical Weber number $We_{\mathrm{crit}}$ is used as 1.1, consistent with measurements from air-water bubbly flow experiments conducted under comparable conditions (Hesketh et al. 1987). As $\mathcal{L}$ increases, the Hinze scale $d_H$ grows linearly on a log-log scale (following power law). Downstream, the dissipation $\varepsilon$ decreases markedly (by approximately 90% over the test section; see Figure 6). Through Equation 4.4, this gives $d_H \propto \varepsilon^{-2/5}$, so a rapid fall in $\varepsilon$ produces only a gradual increase in $d_H$. If $\varepsilon \sim \mathcal{L}^{-m}$ with $m \approx 1.85$ to 2.2 (refer Figure 6), then $d_H \sim \mathcal{L}^{2m/5}$, that is $d_H \sim \mathcal{L}^{0.74}$ to $\mathcal{L}^{0.88}$. Thus, the growth of $d_H$ with $\mathcal{L}$ becomes linear on log-log scale.

| Flow Rate | $\beta_{\exp}$(experimental) | | | | | | $R^2$ | | | | | |
|---|---|---|---|---|---|---|---|---|---|---|---|---|
| | $\phi_1$ | | $\phi_2$ | | $\phi_3$ | | $\phi_1$ | | $\phi_2$ | | $\phi_3$ | |
| $V$(m/s) | $d_{32}$ | $d_{99.8}$ | $d_{32}$ | $d_{99.8}$ | $d_{32}$ | $d_{99.8}$ | $d_{32}$ | $d_{99.8}$ | $d_{32}$ | $d_{99.8}$ | $d_{32}$ | $d_{99.8}$ |
| 6.1 | 0.45 | 0.48 | 0.49 | 0.48 | 0.46 | 0.46 | 0.97 | 0.96 | 0.98 | 0.96 | 0.98 | 0.94 |
| 7.4 | 0.46 | 0.49 | 0.47 | 0.46 | 0.45 | 0.44 | 0.97 | 0.92 | 0.98 | 0.96 | 0.99 | 0.94 |
| 8.4 | 0.46 | 0.53 | 0.45 | 0.47 | 0.47 | 0.50 | 0.99 | 0.95 | 0.98 | 0.95 | 0.98 | 0.94 |

Table 1. Power law coefficient ($\beta_{\exp}$) and coefficient of determination ($R^2$), for the fits for the axial variation of $d_{99.8}$, $d_{32}$, and $d_H$ shown in Figure 10.

In comparison, $d_{99.8}$ also grows linearly (on log-log; general power law) with a smaller slope along the axial direction and remains below $d_H$ at all axial locations except near the inlet ($\mathcal{L} \approx 3.6$). Near the inlet, the size distribution straddles the Hinze scale, so bubbles exist on both sides of $d_H$; those with $d > d_H$ are susceptible to turbulence–induced breakup. Downstream, as turbulence decays, $d_H$ increases faster than $d_{99.8}$, so $d_{99.8}$ falls below and stays below $d_H$. After $\mathcal{L} > 8.2$, roughly no bubbles exceed $d_H$, which indicates a coalescence–dominated regime with active breakup effectively suppressed. The $d_{32}$ also shows a power law growth (linear on log-log) with $\mathcal{L}$ (except near the inlet), but it remains well below both $d_{99.8}$ and $d_H$. The monotonic rise of $d_{32}$ is consistent with coalescence-driven growth.

For $\mathcal{L} \geq 8.2$, both $d_{99.8}$ and $d_{32}$ follow power law trends with $\mathcal{L}$ in Figure 10, and the corresponding fit parameters are listed in Table 1. The fits are strong, with $R^2$ values close to 0.95. The fitted exponents $\beta_{\exp}$ lie within 0.45 to 0.51, indicating that the mean and the upper-tail bubble sizes increase at comparable rates in the coalescence-dominated regime.

The theoretical scaling for the growth of $d_{32}$ and $d_{99.8}$ follows from the effective coalescence rate $\Gamma$, defined as the product of the collision frequency $h$ and the coalescence efficiency $\lambda$ (Coulaloglou & Tavlarides 1977; Prince & Blanch 1990a):

$$\Gamma = h\,\lambda.$$

Classical inertial-range scaling (Kolmogorov 1991), as presented in standard treatments (Batchelor 1953; Monin & Yaglom 1975), underpins collision models for bubbly dispersions. In the inertial subrange of turbulence, the velocity increment $u_{\rm rel}$ across a bubble scale $d$ follows Kolmogorov similarity, resulting in (Batchelor 1953; Monin & Yaglom 1975)

$$u_{\rm rel}(d) \sim (\varepsilon d)^{1/3}.$$

where $\varepsilon$ is the mean turbulent dissipation rate. Combining the inertial–range relative velocity with the geometric cross-section $\sim d^2$ yields the turbulent collision kernel:

$$h \propto u_{\rm rel}(d) \times d^2 \propto \varepsilon^{1/3}\,d^{7/3}.$$

| $\varepsilon$ scaling | $d$ ($d_{32}$ and $d_{99.8}$) scaling |
|---|---|
| $\varepsilon \propto \mathcal{L}^{-m}$ | $d \propto \mathcal{L}^{\beta_{th}}$ |
| $\varepsilon \propto \mathcal{L}^{-1.85}$ | $d \propto \mathcal{L}^{23/40}$ |
| $\varepsilon \propto \mathcal{L}^{-2}$ | $d \propto \mathcal{L}^{1/2}$ |
| $\varepsilon \propto \mathcal{L}^{-2.1}$ | $d \propto \mathcal{L}^{9/20}$ |

Table 2. Bubble diameter scaling in highly decaying turbulent flow. Here $\beta_{th} = \dfrac{3-m}{2}$.

Further simplification results in,

$$\Gamma = 8C\,\lambda\,(\varepsilon)^{1/3}\,d^{\frac{7}{3}}\,. \tag{4.5}$$

The kernel has units of volume per unit time, and the coalescence efficiency satisfies $\lambda \in [0, 1]$. The evolution of the number density $n$ can then be expressed using Smoluchowski's binary coagulation equation (accounting for double-counting and assuming a monodisperse distribution) (Kreer & Penrose 1994):

$$\frac{\mathrm{d}n}{\mathrm{d}t} = -\tfrac{1}{2}\,\Gamma\,n^2. \tag{4.6}$$

For duct flow, substituting $t = \mathcal{L}/V$ and $n = 6\phi/(\pi d^3)$ into Equation 4.6, and simplifying for decaying turbulence with $\varepsilon = \varepsilon(\mathcal{L})$ (refer Appendix C), we get:

$$\frac{\mathrm{d}d}{\mathrm{d}\mathcal{L}} \equiv K[\varepsilon(\mathcal{L})]^{1/3}d^{\frac{1}{3}}, \tag{4.7}$$

where $K$ depends on $\lambda$. Under the present flow conditions, $\lambda$ is evaluated using the commonly employed film drainage model and is found to be a constant (refer Appendix C). Using a power law decay referenced to a finite start location $\mathcal{L} > 0$ (see Equation 4.1):

$$d(\mathcal{L}) \propto \mathcal{L}^{\frac{3-m}{2}} \propto \mathcal{L}^{\beta_{th}}, \quad \text{where } \beta_{th} = \frac{3-m}{2}. \tag{4.8}$$

The theoretical power law coefficient $\beta_{th}$ in Equation 4.8 is listed in Table 2. For $m$ in the range 1.85-2.10, $\beta_{th}$ varies from 0.45-0.57, which closely matches the experimentally fitted exponents for $d_{32}$ and $d_{99.8}$, 0.44 to 0.53, in Table 1. Thus, a steeper dissipation decay (larger $m$) corresponds to a larger $\beta_{th}$. At fixed $m$, higher sustained $\varepsilon$ increases the collision frequency and accelerates bubble-size growth. For applications that require tight control of the size distribution (for example, targeting $d_{32}$ or $d_{99.8}$), one may adjust $m$ by passive means such as modifying wall conditions or by active means such as acoustic or ultrasonic forcing. To make $d$ grow in parallel with the Hinze scale, one could equate the scalings $d \sim \mathcal{L}^{(3-m)/2}$ and $d_H \sim \mathcal{L}^{2m/5}$, which gives $m = 5/3$. At $m = 5/3$, both $d_{32}$ and $d_{99.8}$ remain self-similar and grow in parallel with $d_H$ as $\mathcal{L}$ increases.

Near the inlet, the pure-coalescence scaling does not apply because breakup remains active. Figure 10 shows that, owing to breakup and an initially unstable distribution, $d_{99.8}$ is capped and grows at a different rate than $d_{32}$; consequently, the ratio $\mathcal{D} = d_{99.8}/d_{32}$ evolves through the entry region before saturating downstream (once the two growth curves become roughly parallel). Consistent with this behavior, Razzaque <u>et al.</u> (2003<u>b</u>) reported that monodisperse injections ($\mathcal{D} \approx 1$) in fully developed, coalescence-dominated turbulence relax to a constant $\mathcal{D} \approx 2.2$. Monitoring $\mathcal{D}$ is operationally important in mixing, heat-transfer, and reactor applications because it governs interfacial area and distribution shape (Lehr et al. 2002b; Risso 2018).

### 4.6. *Evolution of bubble size ratio $\mathcal{D}$ along the duct*

The axial evolution of the extreme-to-mean bubble-size ratio $\mathcal{D}$ (= $d_{99.8}/d_{32}$) quantifies tail heaviness relative to the interfacial-area–weighted mean, providing a robust, outlier-resistant proxy for the largest statistically reliable bubbles (preferable to raw $d_{max}$) and for population broadening as turbulence decays (Razzaque et al. 2003a; Risso & Fabre 1998). While increasing $\mathcal{D}$ indicates tail amplification via coalescence, a near-constant $\mathcal{D}$ indicates shape stabilization (self-similarity); tracking $\mathcal{D}(\mathcal{L}, V, \phi)$ thus provides a concise regime marker and a useful calibration target for population-balance models. Applications include: (i) electrochemical reactors (electrolysis, $CO_2$) - lower $\mathcal{D}$ stabilizes mass-transfer layers and overpotential, maintaining a higher, more stable current density, and restricts electrode coverage by infrequent, large bubbles (Taqieddin et al. 2017; Angulo et al. 2020); (ii) gas-lift risers and multiphase pipelines - bounding $\mathcal{D}$ helps maintain bubbly/churn flow, stabilize pressure drop, and lift efficiency, as a rising $\mathcal{D}$ is an early indicator of Taylor-bubble/slug inception (Diaz et al. 2024; De Temmerman et al. 2015); (iii) membrane systems (air-scouring): while large tails decrease wall coverage and speed up fouling, tuning $\mathcal{D}$ maximizes near-wall renewal and scour frequency at fixed aeration power (Alameedy et al. 2025; Fabre & Liné 1992).

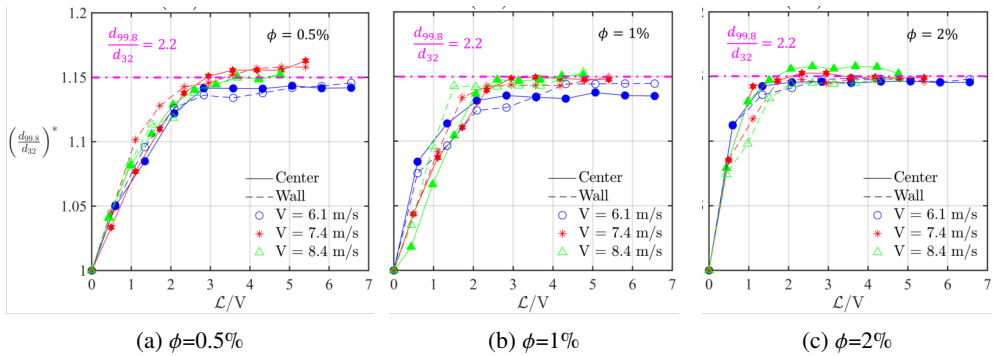

(a) $\phi$=0.5%        (b) $\phi$=1%        (c) $\phi$=2%

Figure 11. Variation of $(\mathcal{D})^*$(normalised) in developing turbulent duct flow.

Figure 11 presents the streamwise evolution of the normalised bubble-size ratio $\mathcal{D}^*$ (normalised by $\mathcal{D}_{\mathcal{L}=0}$) for three bulk void fractions (0.5%, 1% and 2%) at bulk velocities of 6.1, 7.4, and 8.4 m/s. The initial $\mathcal{D}_{\mathcal{L}=0}$ is nearly invariant across $\phi$ and $V$ because intense inlet turbulence rapidly breaks larger bubbles and promotes swift coalescence of smaller ones, yielding a narrow distribution. Normalizing by $\mathcal{D}_{\mathcal{L}=0}$ therefore sets a consistent reference scale and facilitates comparison across cases. Data are shown at two positions (centerline and near-wall - refer Figure 1) to assess spatial uniformity. The figure shows $(\mathcal{D})^*$ is virtually the same at the duct center and near the wall (differences <5%), indicating that the bubble size distribution width is almost uniform across the cross-section. This is consistent with the expectation that in a fully developed (dynamic equilibrium) bubbly flow, the size distribution becomes independent of position (Kleinstreuer & Agarwal 2001). Across all the variables, $(\mathcal{D})^*$ stays within a wide range of approximately 1 to 1.15. In other words, even under varying flow rates and $\phi$, the largest bubbles are only about twice the Sauter mean diameter.

In the developing region (near the inlet), $(\mathcal{D})^*$ rises with axial distance $\mathcal{L}$, reflecting the broadening of the bubble size distribution due to coalescence. Farther downstream, the ratio levels off, approaching an asymptotic value. Higher void fractions (1% and 2%) reach this plateau sooner (small $t = \mathcal{L}/V$), whereas at 0.5% void fraction the ratio takes a longer time (i.e, $\mathcal{L}/V$). This trend indicates a coalescence-dominated flow development – as bubbles travel downstream, frequent coalescence creates larger bubbles and a wider size spread until a balance is achieved. The eventual "saturation" value of $(\mathcal{D})^*$ falls around 1.15 (where $\mathcal{D} \approx 2.2$) for all velocity and void-fraction combinations (with the caveat that the 0.5% case may not have fully reached its plateau in the available length). Once this value is attained, the distribution width no longer grows with $\mathcal{L}$, implying that breakup and coalescence rates have equilibrated.

The ratio $(\mathcal{D})^*$ serves as a quantitative proxy for the width or polydispersity of the bubble size distribution. A value of 1.15 implies that the upper end of the size spectrum (excluding only the extreme outliers) is about 1.15 times the mean size, which points to a moderately broad distribution. Initially, near the inlet, this ratio is lower – the bubble population is relatively narrow, likely because the intense turbulent breakup of the injected air produces predominantly small bubbles with few very large ones. As the flow develops downstream, bubbles interact: smaller bubbles coalesce into larger ones, stretching the upper tail of the size distribution. This causes $d_{99.8}$ (a near-maximum diameter) to grow faster than $d_{32}$, and thus $(\mathcal{D})^*$ increases. Eventually, a transition occurs where further downstream evolution is minimal: the competing processes of bubble coalescence (which broadens the distribution) and bubble breakup or dispersion (which tends to narrow it by capping the maximum size) reach a dynamic balance. At this point, the ratio $(\mathcal{D})^*$ stabilizes at an approximately constant value. The attainment of a constant ratio signifies that the shape of the bubble size distribution has become self-similar (or self-preserving) along the duct. In practical terms, if one rescales the bubble diameters by a characteristic size (e.g. $d_{32}$), the probability distribution of $d$ collapses to the same curve at different axial locations once this equilibrium is reached. In our experiments, this self-similar distribution is essentially achieved for 1% and 2% void fractions (plateau 1.15), whereas the 0.5% case is still approaching that state.

The observed plateau ratio of 1.15 is in agreement with previous studies of fully developed turbulent bubbly flows. Razzaque et al. (2003b) reported that in horizontal pipe flows, the size ratio $\mathcal{D}$ consistently approached 2.2 in coalescence-dominated fully developed turbulent bubbly flows. Notably, they found this ratio to be nearly independent of axial position, flow velocity, and $\phi$ in the coalescence-dominated regime, mirroring our finding that radial position and flow conditions have little influence once the distribution equilibrates. Prince & Blanch's classic bubble column experiments and population-balance model likewise indicated that beyond a certain column height (the "dynamic equilibrium region"), the bubble size distribution becomes independent of location (Prince & Blanch 1990b) – a hallmark of a self-preserving distribution. In such equilibrium, the distribution's width is effectively fixed: Hesketh et al. (1987), for example, conjectured a constant value for the standard deviation of the distribution in fully developed pipe flow. Similarly, Lehr et al. (2002a) observed in their simulations that once coalescence and breakup balance out in a bubble column, the shape of the size distribution no longer changes with axial distance. Lance & Bataille (1991)'s uniform bubbly flow experiments can be seen as a limiting case: by injecting bubbles of nearly uniform size, they achieved a statistically steady (monodisperse) distribution, illustrating the extreme end of the narrow-width scenario. Across these studies, a consensus emerges that a stable width ratio $(\mathcal{D})^* \approx$ 1.15 characterizes the self-similar bubble size distribution in developed turbulent flows. This consistency suggests that $(\mathcal{D})^*$ is a robust metric for comparing distribution breadth across different systems. In summary, the evolution of $(\mathcal{D})^*$ along the duct encapsulates the shift from an initially coalescence-dominated dispersion (yielding a relatively narrow size range) to a pure coalescence growth of large bubbles (broadening the distribution). The plateau in $(\mathcal{D})^*$ provides a convenient quantitative indication that a self-preserving bubble size distribution has been attained. In engineering terms, this single ratio concisely captures the development and eventual stabilization of the bubble population – offering a practical measure of when a bubbly flow has reached a state of geometric similarity in its size distribution, as evidenced by the common asymptotic value of about 2.2 reported in the literature. Such a metric is valuable for comparing flow conditions and for validating models of breakup/coalescence, since any deviation from the expected 2.2 at equilibrium (as observed under certain extreme turbulence conditions) can signal a breakdown of classical self-similarity assumptions (Razzaque et al. 2003b). The present results, in line with prior works, underscore that $(\mathcal{D})^*$ is an insightful parameter for tracking and quantifying the approach to equilibrium in polydisperse bubbly flows.

### 4.7. *Universality in temporal variation of bubble sizes*

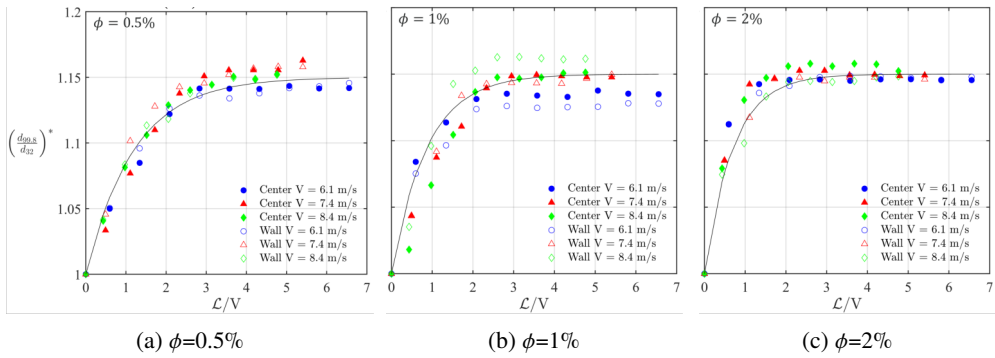

Figure 12. Transient variation of $(\mathcal{D})^*$(curve fitted) in developing turbulent duct flow

As seen in the previous section, the normalised bubble size ratio $(\mathcal{D})^*$ increases monotonically with axial distance. Empirically, the evolution follows an exponential relaxation towards an asymptote as seen in Figure 12 and fitted with

$$\left(\frac{d_{99.8}}{d_{32}}\right)^* = 1.15 - 0.15\,e^{-\frac{t}{\tau}} \; ; \text{ where } \; \left(\frac{d_{99.8}}{d_{32}}\right)^* = \frac{\frac{d_{99.8}}{d_{32}}}{\frac{d_{99.8}}{d_{32}}\,|_{x=0}} \tag{4.9}$$

$$\Delta \left(\frac{d_{99.8}}{d_{32}}\right)^* \% = \frac{\left(\frac{d_{99.8}}{d_{32}}\right)^*_{\max} - \left(\frac{d_{99.8}}{d_{32}}\right)^*}{\left(\frac{d_{99.8}}{d_{32}}\right)^*_{\max}} \times 100 \approx 13\,e^{-\frac{t}{\tau}} \tag{4.10}$$

where $t = \mathcal{L}/V$ is the convective axial coordinate (distance traveled $\mathcal{L}$ divided by bulk velocity $V$) and $\tau$ is a characteristic *relaxation time* (or equivalently, time scale) for coalescence. At the flow inlet ($x = 0$), $(\mathcal{D})^* = 1$ by definition; downstream it rises toward an asymptotic value of 1.15 (a 15% increase in the extreme-to-mean diameter ratio). This exponential trend indicates a first-order process: the rate of change of the size ratio is highest near the inlet (where the distribution is narrow) and decays progressively as the distribution broadens and approaches a new equilibrium state. Such behavior is consistent with classical coagulation kinetics, wherein the population's evolution can be characterized by a single dominant time scale (Ruiz-Rus et al. 2022). The form of the above relation is analogous to Smoluchowski's theory for Brownian coalescence (with $\tau$ playing the role of a *"half-life"* of the initial population) (Ruiz-Rus et al. 2022). Physically, as coalescence proceeds, the diminishing number of bubbles (and the widening size spread) reduces the net collision rate, naturally yielding an exponential approach to a steady state.

Figure 12 shows that the axial evolution of the normalized diameter ratio, $(\mathcal{D})^*$, as a function of residence time ($t$), demonstrates a universal trend. Symbols represent experimental data measured at different bulk velocities ($V = 6.1$–$8.4$ m/s) and radial positions (centerline and near-wall) for initial void fractions of $\phi = 0.5\%$, $1\%$, and $2\%$ in Figure 12. All data collapse onto a single master curve for a given $\phi$. The curve in each panel is the best-fit exponential equation 4.9, illustrating the rapid rise from unity at $t = 0$ toward an asymptote of about 1.15. Higher void fraction yields a steeper rise (smaller $\tau$), consistent with more frequent bubble coalescence at larger $\phi$.

The $(\mathcal{D})^*$ plotted versus time $t$, for all $V$ measurements - near-wall and centerline — collapse onto a single normalized curve at fixed initial $\phi$ (see Figure 12), implying a degree of universality in the coalescence dynamics. Consistently, the center–wall distribution ratio traces an indistinguishable trajectory in $(\mathcal{D})^*$ vs. $t$ space. In other words, the evolution of $(\mathcal{D})^*$ is set chiefly by the bubble-population (i.e., $\phi$) and $t$, and independent of radial positions. The influence of different bulk velocities is absorbed by the $t = \mathcal{L}/V$ scaling – faster flows simply convect bubbles further in a given time, but the coalescence progression per unit time is unchanged. Such behavior suggests that the coalescence process is governed by the cumulative interaction time between bubbles, and that our normalization captures the essential time scale of those interactions. Similar universal behavior has been reported in controlled coalescing swarms: for example, Ruiz-Rus et al. (2022) found that the growth of a characteristic large-bubble diameter ($D_{V90}$, 90th percentile of bubble diameter) downstream of an injector could be collapsed to a single curve when distances were non-dimensionalized, indicating a common coalescence "cascade" process independent of injection rate. In our data, the master curve is well-described by the above exponential fit, reinforcing the notion that a first-order kinetics governs the axial broadening of the bubble size distribution.

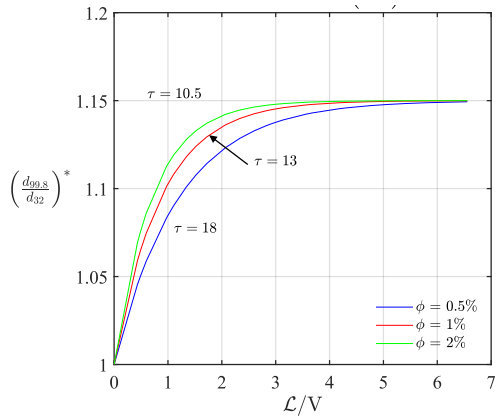

Figure 13. Transient variation of $(\mathcal{D})^*$ and its deviation from equilibrium with residence time

Each master curve is, however, parameterized by the initial void fraction $\phi$. The relaxation length (or e-folding length) $\tau$ in the exponential fit depends strongly on $\phi$, whereas other parameters (e.g. bulk velocity, radial position) have negligible influence on $\tau$. We observe that higher void fractions yield smaller $\tau$, i.e., a more rapid approach to the asymptote. For instance, at $\phi = 2\%$ the diameter ratio rises quickly and plateaus within a short time (Figure 13), whereas at $\phi = 0.5\%$ the rise is much more gradual. Quantitatively, increasing $\phi$ from 0.5% to 2% reduces $\tau$ by roughly a factor of 18 to 10.8 (as inferred from the curves in Figure 13), indicating that bubble populations with more initially crowded conditions coalesce and relax to their new size distribution much sooner. This trend aligns with physical intuition: a higher $\phi$ means a higher number density of bubbles and more frequent bubble–bubble encounters. As a result, coalescence events occur more frequently, accelerating the growth of larger bubbles and broadening of the distribution. In contrast, at lower void fraction, bubbles are widely separated and may travel considerable distances before encountering a partner to coalesce with (Ruiz-Rus et al. 2022). Our findings mirror those of other studies in which void fraction was shown to be a key parameter controlling coalescence rates. For example, in the experiments of Riviere et al. (2022), coalescence in a confined vertical swarm did not even commence until a certain downstream distance for $\phi < 3\%$, whereas at $\phi \approx 5\%$ coalescence began almost immediately near the injection point. This indicates that $\tau$ is essentially an increasing function of the initial bubble spacing (or inverse function of $\phi$). Importantly, $\tau$ in our flow appears to be independent of the mean velocity $V$ — consistent with the notion that, in the reference frame of the flow, the intrinsic coalescence time scale is set by bubble collision kinetics (dictated by $\phi$ and turbulence levels) rather than convective transport.

The exponential broadening of the size ratio can be attributed to several coalescence mechanisms acting in tandem. First, random turbulent fluctuations in the liquid velocity field induce stochastic bubble–bubble encounters. Turbulence can bring bubbles together from different streamlines, effectively increasing the collision frequency beyond what pure laminar drift would produce (Prince & Blanch 1990b). In our high Re flow, both the background turbulence and bubble-induced agitation contribute to such random collisions. Turbulence-induced coalescence has been modeled as a dominant mechanism in similar systems (e.g. Bröder & Sommerfeld (2004) and Kamp et al. (2001) developed a mechanistic model for bubble coalescence driven by turbulent eddies). Second, velocity shear and differential buoyancy across the flow contribute to collisions. Bubbles rising in a shear flow experience different velocities; a faster-rising bubble may overtake a slower bubble ahead, leading to a collision from behind. Likewise, velocity gradients near the wall can push bubbles together. Prince & Blanch's classic analysis of bubble column coalescence considered collisions arising from turbulence, buoyancy-induced velocity differences, and laminar shear as the three primary sources of inter-bubble collisions (Prince & Blanch 1990b). All three are relevant here: even in a co-current flow, larger bubbles have higher rise velocities relative to the liquid and can catch up with smaller ones (a buoyancy-driven overtaking collision), while shear in the liquid (especially near walls or in velocity profiles) can create lateral motion that brings bubbles into contact (Prince & Blanch 1990b). Third, radial migration effects in the flow can enhance coalescence among certain bubble groups. In a vertical shear flow, small nearly spherical bubbles tend to experience a lift force directing them toward the wall, whereas larger deformable bubbles experience lift towards the center of the pipe (Lucas et al. 2023). This size-dependent lateral segregation means that as bubbles coalesce and grow, they preferentially accumulate in the core region. The core thus becomes enriched in large bubbles, which increases the local collision rate among these large bubbles. Meanwhile, smaller bubbles concentrate near the periphery, where their lower abundance (and possibly lower relative velocities) results in fewer collisions among themselves. The net effect of this radial rearrangement is to hasten the growth of the largest bubbles (by concentrating them together) and, conversely, to decouple the smaller bubbles from frequent encounters with the largest ones. In addition, as large bubbles rise, they generate wake regions that can entrain following bubbles – a mechanism that can draw smaller bubbles into collisions from behind. High-speed visualizations of coalescing swarms indeed show bubble pairs forming via wake capture and subsequent coalescence in the bubble wake's low-pressure region (Ruiz-Rus et al. 2022). The radial bubble distribution has been widely studied and arises from the concurrent action of turbulent dispersion, shear-induced collisions, lateral lift–induced migration, wake entrainment, and differential rise due to buoyancy (negligible here) (Tomiyama 1998; Tomiyama et al. 2002). Their combined outcome is a rapid initial flurry of coalescence events (when many small bubbles are present in close proximity), followed by a slower rate of coalescence as the bubble count decreases and an equilibrium size distribution is approached.

It is worth noting that the asymptotic normalized ratio of 1.15 observed here corresponds to a new quasi-steady distribution that is broader than the inlet distribution but not unbounded in width. In the present coalescence-dominated conditions, bubble breakup is negligible in the measurement range, so the plateau in $(\mathcal{D})^*$ presumably represents a balance where further coalescence is limited by the decreasing collision frequency (and possibly by short residence time in the test section). In other systems, if bubbles continue to coalesce unchecked over longer distances, one might expect $(\mathcal{D})^*$ to continue rising slightly above 1.15. However, physical constraints often intervene to cap the distribution width. For example, extremely large bubbles become susceptible to deformation and *breakup* in strong turbulent flows. Ruiz-Rus et al. (2022) observed that at sufficiently downstream locations (or at very high $\phi_0$), the largest bubbles in their confined swarm began to break apart, counteracting further growth of the diameter extremes. Thus, an ultimate steady state in a very long pipe or high void fraction scenario would be reached when coalescence and breakup equilibrate, yielding a stabilized size distribution. In bubble column experiments, it has indeed been observed that beyond a certain distance (on the order of a few column diameters), the bubble size distribution ceases to evolve significantly, indicating the onset of a statistically steady state (Jo & Revankar 2010). In our case, the flow is *coalescence-dominated* and the test section length is such that an asymptotic state is nearly attained (coalescence rates have slowed appreciably by the end of the test section), but breakup has not yet become significant. The modest asymptotic increase of 15% in the $(\mathcal{D})$ ratio thereby reflects the extent of coalescence achievable before the system either runs out of opportunities (limited residence time) or enters a new regime where other processes (breakup, mass transfer, etc.) intervene.

From a *modeling perspective*, these findings provide quantitative guidance for predicting bubble size distribution evolution in coalescence-dominated flows. The fact that the data collapse onto a single exponential curve for each $\phi$ suggests that one can describe the coalescence-induced growth with a simple first-order ODE or closure relation. For example, one may write an evolution equation for the diameter ratio (or for any other measure of distribution width) of the form

$$\frac{d}{dt}\left(\frac{d_{99.8}}{d_{32}}\right)^* = \frac{K}{\tau}\left[\left(\frac{d_{99.8}}{d_{32}}\right)^*_{t=0} - \left(\frac{d_{99.8}}{d_{32}}\right)^*\right] \tag{4.11}$$

with,

$$\left(\frac{d_{99.8}}{d_{32}}\right)^*_{t=0} = \left(\frac{\frac{d_{99.8}}{d_{32}}|_{\text{equilibrium}} \approx 2.2}{\frac{d_{99.8}}{d_{32}}|_{\text{initial}}}\right) \tag{4.12}$$

where $K \approx 1$ is a constant and $\tau = \tau(\phi)$. This would integrate to the observed exponential profile. More fundamentally, the *population balance equation* (PBE) for the bubble size distribution could be informed by our results. The PBE (a generalized Smoluchowski equation) uses coalescence kernels to describe the rate at which bubbles of given sizes collide and merge (Ruiz-Rus et al. 2022). Our observation of an exponential approach to a self-similar state indicates that a *coalescence kernel dominated by binary interactions and a roughly size-independent coalescence efficiency* can reproduce the data. It also suggests that the coalescence kernel (or at least its dominant eigenmode) leads to an approximately exponential decay of bubble count and growth of large-bubble fraction with distance. Such insights can be used to validate and tune coalescence models in CFD simulations. Recent studies have indeed emphasized the need for accurate coalescence closures in Euler–Euler simulations of poly-dispersed flows (Lucas et al. 2023). The present results indicate that a coalescence time scale can be formulated as a function of local void fraction alone, $\tau = \tau(\phi)$, simplifying the closure of coalescence source terms in two-fluid models. For instance, interfacial area transport equations (Ruiz-Rus et al. 2022), which track the evolution of the gas–liquid interfacial area density, include sink terms for area reduction due to coalescence. The dependence of $\tau$ on $\phi$ (faster coalescence at higher void fraction) corroborates the form of many coalescence kernels used in literature, where collision frequency is proportional to the square of bubble concentration (Ruiz-Rus et al. 2022). Modelers can incorporate an empirical correlation for $\tau(\phi)$ gleaned from our data to improve predictions of bubble size distribution downstream of injection. Finally, the *universality* of the normalized coalescence curve implies that, once calibrated for a given $\phi$, the model can be expected to hold across a range of flow velocities and radial positions, at least in the developing bubbly flow regime where buoyancy and inertia are in balance. This is encouraging for the development of robust one-dimensional models for bubble population evolution, as well as for multi-group population balance frameworks like MUSIG, since it suggests that complex flows can be reduced to a simple coalescence law with a single key parameter ($\phi$) capturing most of the variability (Kamp et al. 2001). In summary, the axial evolution of $(\mathcal{D})^*$ in our experiments embodies the essential physics of coalescence-dominated bubbly flows – rapid initial broadening due to frequent collisions, followed by an asymptotic approach to a new equilibrium – and provides both physical insights and quantitative benchmarks for advanced modeling efforts in multiphase flow dynamics.

### 4.8. *Effect of decaying turbulent flow on local void fraction*

In order to explore the radial variation of the local void fraction, the observation window was divided into colored regions (refer right Figure 2), corresponding to near-wall (red) and central (blue) zones. For each region, several hundred 2D images were analyzed to extract the bubble size distribution, repeating until the results became statistically stationary. The local void fraction in each region was then determined as the volume-weighted mean bubble volume per image, normalized by the total region volume. The local void fraction is denoted by $\phi_{wall}$ for near-wall and $\phi_{center}$ for centerline. While these averages are assumed uniform within each colored zone, minor radial variations may exist (Du Cluzeau et al. 2019; Lu & Tryggvason 2006; Bunner & Tryggvason 2002, 2003; Balachandar & Eaton 2010).

Figure 14 shows the change in local void fraction $\Delta\phi = \frac{\phi_{center/wall} - \phi_0}{\phi_0}$ at center and wall in axial direction of the duct across three $\phi$ (0.5%, 1%, 2%). At the inlet ($\mathcal{L} = 0$), the void fraction is nearly uniform, with both centerline and near-wall values equal to the cross-sectional average (input value). As the flow develops downstream, a distinct segregation emerges: the near-wall void fraction decreases steadily, while the centerline void fraction increases commensurately, indicating progressive bubble accumulation in the core. For the lowest $\phi$ of 0.5%, this redistribution is gradual but substantial; by the end of the measured domain, the centerline void fraction rises to $\sim 1.6\times$ its inlet value, while the near-wall void fraction drops to $\sim 0.4\times$ its inlet value, resulting in a growing centre-wall void fraction difference ($\Delta\phi$) that increases nearly linearly with time and shows no plateau. At higher $\phi$ ($= 2\%$), a similar trend is observed with larger absolute changes: the centerline void fraction rises from 2% to nearly 3%, while the near-wall region is depleted to about 1%. However, when normalized by the initial average, the relative enrichment at the core is slightly less pronounced at higher $\phi$, indicating that bubble–bubble interactions and enhanced mixing at high void fractions can temper the relative segregation (Bunner & Tryggvason 2002, 2003; Dabiri et al. 2013; Burns et al. 2004b).

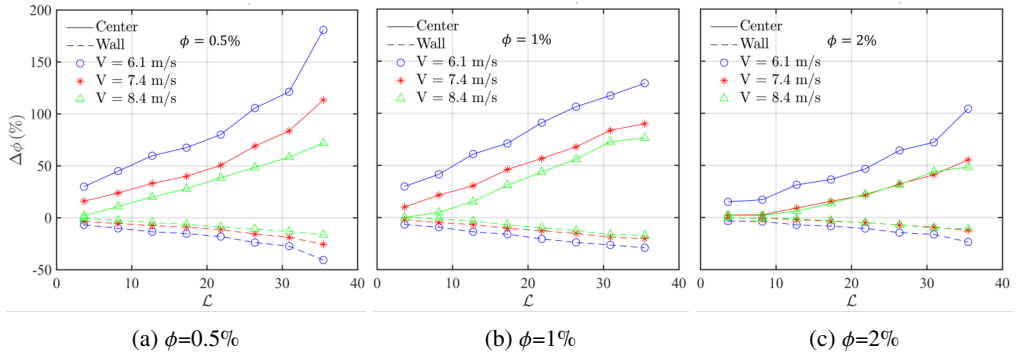

(a) $\phi$=0.5%　　　　　　(b) $\phi$=1%　　　　　　(c) $\phi$=2%

Figure 14. Variation of $\phi_{wall}$ and $\phi_{center}$ with axial position ($\mathcal{L}$).

The influence of mean flow velocity is also evident. At lower bulk velocities ($V \approx$ 6.1 m/s), the transition toward a core-peaked profile occurs more rapidly and distinctly, with bubbles segregating inward sooner and to a greater extent. At higher velocities ($V \approx$ 8.4 m/s), void fraction profiles remain closer to uniform for longer axial distances, and the center-wall divergence grows more slowly. This arises because stronger turbulence at higher $V$ is sustained farther downstream, keeping bubbles smaller and better mixed, and thus delaying the onset of segregation (Lu & Tryggvason 2006; Balachandar & Eaton 2010; Burns et al. 2004b). At lower $V$, turbulence decays more quickly, enabling earlier bubble coalescence and growth, which amplifies the effect of lift and buoyancy on radial migration (Bunner & Tryggvason 2003; Lu & Tryggvason 2006; Balachandar & Eaton 2010).

These observations represent a significant departure from classical duct/pipe flow studies, which consistently report wall-peaked void fraction profiles in turbulent flows (Hibiki & Ishii 1999; Delnoij et al. 1997). In these flows, small bubbles - owing to positive lift coefficients - accumulate near the wall, producing pronounced wall peaks. For our flow, however, even with small bubble sizes ($d_{32} \sim 200–600~\mu m$), a strong and persistent core peak emerges. This divergence from established patterns indicates that bubble migration mechanisms in high Reynolds number, vertical bubbly flows are fundamentally distinct. A key novelty of the present study is the demonstration that core-peaking arises for bubbles an order of magnitude smaller than the classical lift-reversal diameter (typically 5–6 mm in air–water systems) (Tomiyama et al. 2002; Hidman et al. 2022). Traditional lift models predict that only sufficiently large, deformable bubbles undergo a reversal in the lateral lift force, driving them from the wall toward the core and generating center-peaked profiles. Our experiments, however, show that even micron-scale bubbles experience strong inward migration under diminished turbulence, challenging established predictions and aligning with recent DNS results that reveal early lift reversal in deformable bubbles at lower Reynolds numbers (Tomiyama 1998; Hidman et al. 2022; Du Cluzeau et al. 2019). Although Tomiyama's correlation (Tomiyama 1998) for lift coefficient $C_L$ given below (see Equation 4.13) is based on the motion of isolated bubbles at moderate Re and has not been validated for strongly turbulent flows, it does suggest a lift coefficient reversal:

$$C_L = 0.00105\,\text{Eo}^3 - 0.0159\,\text{Eo}^2 - 0.0204\,\text{Eo} + 0.474, \tag{4.13}$$

where $\text{Eo} = \frac{(\rho_\ell - \rho_g)g d_b^2}{\gamma}$ is the Eötvös number, which is defined as ratio of buoyancy to surface forces. Here, $\rho_\ell$ is the liquid density, $\rho_g$ is the gas (air) density, $d_b$ is the bubble diameter, and $g$ is gravitational acceleration. For $d_{32} = 200$–$600\,\mu m$, $\text{Eo} = 0.055$–$0.49$, yielding $C_L \approx 0.47$ - positive and nearly constant - indicating a strong inward-directed lift even at the microscale. Classically, core-peaking is not expected for such small bubbles. For instance, Lu & Tryggvason (2006) found that nearly spherical bubbles remain wall-peaked downstream in highly turbulent flows, and that a lift reversal - leading to core accumulation - requires $\text{Eo} \gg 1$ (i.e., significantly deformable bubbles). However, their study does not address cases where $\text{Eo} \ll 1$, particularly in highly decaying flows, where turbulent dispersion has weakened to the extent that even a modest positive $C_L$ drives bubble accumulation toward the core. As in this case, turbulence decays by over 90% along the duct (see the monotonic drop in $k/V^2$ – refer Figure 5), and turbulent dispersion is dramatically weakened, allowing lift and wall forces to dominate and drive bubbles toward the core (Lu & Tryggvason 2006; Burns et al. 2004a). The redistribution of void fraction can also be described by the drift-flux framework,

$$j_g = C_0 j_m + V_{gj}\phi, \tag{4.14}$$

where $j_g$ is the gas flux, $j_m$ is the mean mixture flux, $V_{gj}$ is the drift velocity, and $C_0$ is the distribution parameter. Here, $C_0 > 1$ (typically $C_0 \approx 1.2$), reflecting the observed strong central enrichment, in contrast to $C_0 < 1$ for wall-peaked profiles in upward flows (Kawaji 2017). Turbulent dispersion acts as a diffusive flux opposing steep void fraction gradients, modeled as

$$\mathbf{M}_{g,\text{disp}} = -\frac{3}{4}\,(1 - \phi)\rho_\ell C_D d_b |U_R|\, D_{\text{eff}}\,\nabla\phi, \tag{4.15}$$

where $\mathbf{M}_{g,\text{disp}}$ is the turbulent dispersion momentum flux of the gas phase, $C_D$ is the bubble drag coefficient, $|U_R|$ is the slip velocity magnitude between gas and liquid, $D_{\text{eff}}$ is the effective turbulent diffusivity, and $\nabla\phi$ is the gradient of the gas volume fraction. As $D_{\text{eff}}$ decreases, the dispersion flux becomes too weak to counteract inward lift and wall-lubrication forces, enabling sharper and more persistent core peaks (Hosokawa & Tomiyama 2009; Ooms et al. 2007; Burns et al. 2004a; Elghobashi 2019; Balachandar & Eaton 2010). The wall-lubrication force further ensures $\phi \to 0$ at the wall, in agreement with analytical models (Marfaing et al. 2016).

The observed transition from uniform to core-peaked void fraction profiles arises from three key mechanisms: (1) as turbulence intensity diminishes (by $\sim$ 90%), turbulent dispersion is weakened, losing its ability to homogenize the bubble distribution and allowing other forces to dominate; (2) even micron-scale bubbles experience sufficient inward lift under low turbulence, overturning the conventional view that a critical bubble size ($\sim$5–6 mm) is necessary for lift reversal and core-peaking, as recently confirmed by DNS (Hidman et al. 2022; Du Cluzeau et al. 2019); and (3) as lift reversal takes hold, bubbles migrate toward the duct centerline, where upward liquid velocity is highest and drag is minimized, energetically favoring central accumulation (Lu & Tryggvason 2006). This drag-reduction mechanism is further reinforced by wall-lubrication, which prevents bubble accumulation near the wall and maintains a depleted boundary region.

These findings not only depart from established wall-peaked paradigms seen in prior studies, but also suggest that the interplay of lift, wall-lubrication, and turbulent-dispersion forces governs radial void fraction profiles even under extreme conditions. The present results thus fill an important gap in the literature, demonstrating persistent and robust core-peaking even for micron-scale bubbles — a regime not previously reported — and provide new insight into dispersed phase redistribution mechanisms in turbulent multiphase flows.

In particular, the monotonic decrease of $\phi$ near the wall and its growth near the center with increasing $\mathcal{L}$ raise important questions about the mechanisms driving migration from the wall toward the center. This behavior warrants further investigation by examining how radial distributions evolve under varying flow conditions and system parameters.

### 4.8.1. *Transient void fraction ($\phi$) evolution*

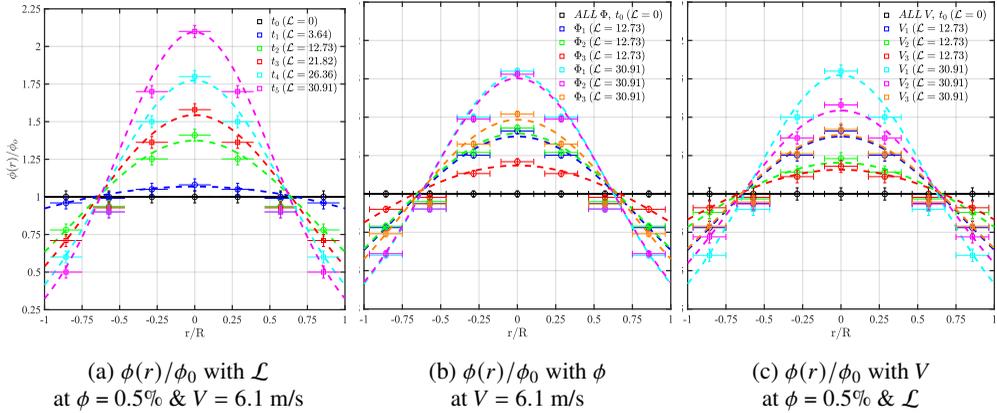

(a) $\phi(r)/\phi_0$ with $\mathcal{L}$ at $\phi = 0.5\%$ & $V = 6.1$ m/s

(b) $\phi(r)/\phi_0$ with $\phi$ at $V = 6.1$ m/s

(c) $\phi(r)/\phi_0$ with $V$ at $\phi = 0.5\%$ & $\mathcal{L}$

Figure 15. Transient radial variation of void fraction ($\phi(r)$) in decaying turbulent regime. **Parameters:** Sampling instants $t_i$ are defined by $t_i = \mathcal{L}_i/V_1$ ($i = 0, \ldots, 5$). Bulk velocities: $V_1 = 6.1$, $V_2 = 7.4$, $V_3 = 8.4$ m/s. Axial locations: $\mathcal{L}_0 = 0$, $\mathcal{L}_1 = 3.64$, $\mathcal{L}_2 = 12.73$, $\mathcal{L}_3 = 21.82$, $\mathcal{L}_4 = 26.36$, $\mathcal{L}_5 = 30.91$.

As bubbles migrate radially from the wall region (red) toward the duct center (blue) (refer Figure 2(b)), the available cross-sectional area decreases. This geometric constraint leads to an amplification of local $\phi$ near the center. Consequently, the radial variation of the local void fraction is inherently non-linear. To quantitatively characterize this evolution, the measured radial void fraction profiles at each axial location ($\mathcal{L}$), void fraction ($\phi$), and bulk velocity ($V$) were fitted to a Gaussian function (shown in Figure 15):

$$\frac{\phi(r)}{\phi_o} = \Phi_i \exp\left(-\frac{(r/R - R_o)^2}{2\sigma^2}\right) \tag{4.16}$$

where $\phi_0$ is the cross-sectional mean, $\Phi_i$ is the peak amplitude (centerline void fraction), $R_o$ is the mean position (fixed at zero by symmetry), and $\sigma$ is the standard deviation describing the distribution width. The fitted parameters of Figure 15 for all test cases are given in Table 3. Figure 15 shows that near the duct inlet ($\mathcal{L} = 0$), the void fraction profile is nearly flat ($\Phi_i = 1$, $\sigma \to \infty$), confirming uniform bubble dispersion. Downstream, as $\mathcal{L}$ increases, $\Phi_i$ rises and $\sigma$ decreases monotonically - indicating enhanced centerline

| $\mathcal{L}$ | $\phi$ | $\Phi_i$ | $\sigma$ |
|---|---|---|---|
| 0 | 0.5% | 1.00 | $\infty$ |
| 2.6 | 0.5% | 1.07 | 1.85 |
| 12.7 | 0.5% | 1.37 | 0.80 |
| 21.8 | 0.5% | 1.54 | 0.69 |
| 26.4 | 0.5% | 1.77 | 0.59 |
| 30.9 | 0.5% | 2.09 | 0.51 |
| - | - | - | - |

(a) $\phi(r)$ with $\mathcal{L}$ at $\phi = 0.5\%$ & V=6.1 m/s

| $\mathcal{L}$ | $\phi$ | $\Phi_i$ | $\sigma$ |
|---|---|---|---|
| 0 | all $\phi$ | 1.00 | $\infty$ |
| 12.7 | 0.5% | 1.37 | 0.80 |
| 12.7 | 1% | 1.39 | 0.80 |
| 12.7 | 2% | 1.18 | 1.15 |
| 30.9 | 0.5% | 1.77 | 0.59 |
| 30.9 | 1% | 1.75 | 0.59 |
| 30.9 | 2% | 1.48 | 0.73 |

(b) $\phi(r)$ with $\phi = 0.5\% - 2\%$ at V=6.1 m/s

| $\mathcal{L}$ | $V_i$(m/s) | $\Phi_i$ | $\sigma$ |
|---|---|---|---|
| 0 | all $V$ | 1.0 | $\infty$ |
| 12.7 | 6.1 | 1.37 | 0.80 |
| 12.7 | 7.4 | 1.20 | 1.08 |
| 12.7 | 8.4 | 1.16 | 1.23 |
| 30.9 | 6.1 | 1.77 | 0.59 |
| 30.9 | 7.4 | 1.54 | 0.69 |
| 30.9 | 8.4 | 1.38 | 0.81 |

(c) $\phi(r)$ with $V$ at $\phi = 0.5\%$ & $\mathcal{L}$

Table 3. Fitted parameters (refer Figure 15) of the normalized radial void fraction profiles for three different cases.

peaking and profile narrowing due to turbulence decay, increased migration time, and flow development. These trends are most pronounced at lower $\phi$ and $V$, as shown in Figure 15(a).

It is interesting to note the effects of $\phi$ and $V$ (Figure 15(b, c)). For moderate $\phi$ ($\leq 1\%$), $\Phi_i$ and $\sigma$ remain nearly unchanged with increasing $\phi$, indicating self-similar profile shapes. However, at $\phi = 2\%$, the distribution broadens ($\sigma$ increases) and the centerline amplitude decreases ($\Phi_i$ drops). This broadening arises from increased bubble–bubble interactions and, crucially, from stronger turbulent dispersion, which depends directly on the radial gradient $\Delta\phi(r)$. According to Equation 4.15, the dispersion flux is proportional to $-\nabla\phi_g$; so larger $\Delta\phi(r)$ (i.e., greater radial contrast) enhances outward mixing and inhibits further accumulation at the center. This self-regulating mechanism — where increased peaking amplifies dispersion — prevents unlimited core enrichment and is rarely captured in classical drift-flux models (Burns et al. 2004a; Hosokawa & Tomiyama 2009; Ooms et al. 2007). A few recent DNS studies further confirm that bubble-driven dispersion follows Kolmogorov-like scaling and is modulated by radial gradients in void fraction (Elghobashi 2019; Balachandar & Eaton 2010). Similarly, increasing $V$ at fixed $\phi$ leads to lower $\Phi_i$ and higher $\sigma$, consistent with higher dissipation, stronger turbulent mixing, and reduced migration time. Thus, broader, less-peaked distributions at high $\phi$ or $V$ reflect the dominance of dispersion and bubble interactions over migration (Tomiyama et al. 2002; Du Cluzeau et al. 2019; Hidman et al. 2022).

The Gaussian parameterization offers more than a compact description: it provides a direct link between observable profile shape, and the balance between migration and dispersion mechanisms. The fitted $\sigma$ quantifies the effectiveness of turbulent dispersion for given flow conditions, while $\Phi_i$ encodes the extent of core accumulation. These parameters are directly relevant for calibrating and validating turbulence-dispersed two-phase flow models (including gradient based closures) (Burns et al. 2004a) and enable systematic comparison with DNS, RANS, or LES. By relating $\sigma$ to turbulence intensity and $\Delta\phi(r)$, this approach enables predictive scaling and cross-regime analysis — bridging detailed experiments, numerical simulations, and practical engineering models.

## 5. Conclusions and future work

We studied an extreme, previously unexplored regime of developing bubbly flow: a high Re ($\approx 1.3 \times 10^5$) duct where pump-driven turbulence is initially intense ($I \geq 30\%$) and decays by $\sim 90\%$ downstream. This rapid decay shifts the interfacial dynamics from early fragmentation to sustained coalescence, driving a systematic evolution of the bubble-size statistics and the void-fraction field. The measurements provide a spatially resolved benchmark of how turbulence decays, coalescence kinetics, and lateral migration jointly shaping the population.

**Key findings:**

- *Turbulence decay:* In highly turbulent bubbly flow, the dissipation decreases with axial distance as a power law, $\varepsilon \sim \mathcal{L}^{-2}$ - slightly slower than, but close to, the canonical HIT scaling ($\sim \mathcal{L}^{-2.2}$).

- *Power law scaling of the pdf:* Near the duct inlet a mixed tail appears with $d^{-3/2}$ for *sub-Hinze* and a steep scaling of $\sim d^{-10/3}$ for *super-Hinze*; farther downstream the pdf falls below *sub-Hinze* and collapses to a single $d^{-3/2}$ scaling.

- *Regime transition with distance:* Near the inlet, vigorous turbulence promotes fragmentation; as $\varepsilon$ drops, a pure coalescence region is observed.

- *Breakup–coalescence regime shift:* Near the inlet, the bubble distribution straddles the Hinze scale, with turbulence-induced breakup limiting the largest sizes; farther downstream, as $\varepsilon$ decays, both $d_{32}$ and $d_{99.8}$ grow sublinearly and remain below $d_{\mathrm{H}}$, confirming a coalescence-dominated regime.

- *Kinetic growth scaling:* Theory and measurements show that $d_{99.8}$ and $d_{32}$ both grow as $\sim \mathcal{L}^{0.5}$ in this decaying turbulent flow, while the Hinze scale (breakup threshold) grows faster, $d_{\mathrm{H}} \sim \mathcal{L}^{0.8}$. Because bubble sizes lag behind the rising breakup limit, breakup weakens and the evolution trends toward a pure coalescence regime.

- *Growth of polydispersity and self-similarity:* The extreme-to-mean ratio $\mathcal{D}$ rises from $\sim 1.9$ near the inlet to a universal plateau $\approx 2.2$ in the developed region, evidencing a self-similar, quasi-equilibrium spectrum.

- *Cumulative spectrum:* The cumulative distribution of $d/d_{32}$ attains a linear log–log slope of $\approx 1.3$ only after $\mathcal{D}$ saturates and the size distribution stabilizes, with earlier onset at larger $\phi$.

- *Breakup ceiling vs. observed maximum size:* Although the Hinze limit grows rapidly with decay ($d_{\mathrm{H}} \propto \varepsilon^{-2/5}$), the measured upper size increases more slowly; $d_{99.8} < d_{\mathrm{H}}$ at all $x/D$, implying breakup never vanishes—bubbles that exceed $d_{\mathrm{H}}$ are quickly fragmented.

- *Radial void-fraction restructuring:* Void fraction evolves from a near-uniform profile to a sharply core-peaked, Gaussian-like distribution that narrows with $x/D$; the centerline $\phi$ rises while near-wall $\phi$ depletes.

- *Lift-force reversal in intense turbulence:* Sustained center peaking — despite small $d_{32}$ — is consistent with early lift-force reversal predicted in high-turbulence regimes, with wall lubrication further maintaining near-wall depletion.

- *Role of $\phi$ and Re:* Increasing $\phi$ accelerates coalescence and shortens the relaxation length toward the universal spectrum; higher Re (bulk $V$) sustains breakup longer and tends to flatten radial segregation, yet the net trend remains inward focusing as turbulence decays.

Overall, these results establish the coalescence-dominated, high-Re developing regime as a distinct operating window where a universal, *sub-Hinze* $d^{-3/2}$ spectrum and a core-peaked void profile emerge concomitantly with turbulence decay, filling a key gap in multiphase-flow understanding and scale-up.

**Future work:**

- Extending the parameter space to higher void fractions and longer ducts will clarify transition boundaries to dense/slugging regimes and test the universality of the $\mathcal{D} \approx 2.2$ plateau.
- The The local void fraction peaks at the centerline (a reversal of the usual near-wall maximum). Current models lack validation in this regime; focused experiments and DNS are required to resolve the mechanism.

**Author contributions**

**Vivek Kumar**: Conceptualization, Methodology, Investigation, Data Curation, Formal Analysis, Visualization, Writing – Original Draft. **Prasoon Suchandra**: Data Curation and Analysis, Methodology, Writing – Review & Editing. **Ardalan Javadi**: Conceptualization, Visualization, Writing – Review & Editing. **Suhas S. Jain**: Supervision, Writing – Review & Editing. **Cyrus Aidun**: Conceptualization, Supervision, Funding Acquisition, Project Administration, Writing – Review & Editing.

**Data Availability**

Raw data are available upon request to the Editor or the corresponding author.

**Acknowledgments**

The information, data, or work presented herein was funded in part by the Advanced Research Projects Agency-Energy (ARPA-E), U.S. Department of Energy, under Award Number DE-AR0001587. The views and opinions of authors expressed herein do not necessarily state or reflect those of the United States Government or any agency thereof. We thank Mr. Jacob Stancil and Mr. Jason Rom for their contributions. The authors acknowledge the use of AI-based tools (e.g. Writefull, Grammarly, ChatGPT) for spelling, grammar correction, and paraphrasing; all technical content, analysis, and conclusions are the authors' own.

**List of Acronyms/Symbols**

| Symbol | Description |
|---|---|
| **Velocity and turbulence** | |
| $u, v, w$ | Velocity components in $x$, $y$, $z$ directions |
| $\overline{\mathbf{u}}$ | Mean velocity vector |
| $u', v', w'$ | Velocity fluctuation components in $x$, $y$, $z$ directions |
| $u_\tau$ | Friction velocity |
| $\overline{u_i'u_j'}$ | Reynolds stress tensor |
| $\mathcal{I}, TI$ | Turbulence intensity |
| $k$, TKE | Turbulent kinetic energy |
| $\varepsilon$ | Turbulence dissipation rate |
| $\lambda$ | Taylor microscale |
| **Physical properties** | |
| $\nu$ and $\gamma$ | Kinematic viscosity and Surface tension |
| $\rho_l, \rho_g$ | Liquid and gas density |
| **Geometry and setup** | |
| $y$ | Distance from the wall (wall-normal coordinate) |
| $x, L$ | Axial position; normalized axial position |
| $D$ | Duct hydraulic diameter |
| $x/D = \mathcal{L}$ | Normalized duct length |
| $t = \mathcal{L}/V$ | Residence time in the duct |
| $b, l, \delta$ | Channel width, axial and lateral positions |
| $R_o$ | mean position of Gaussian Distribution |
| **Flow parameters** | |
| $Q$ | Volumetric flow rate |
| $V = V_1, V_2, V_3$ | Bulk velocity at 72 LPM, 87 LPM and 98 LPM |
| $\phi = \phi_1, \phi_2, \phi_3$ | Void fraction of 0.5 %, 1 % and 2 % |
| **Bubble metrics and statistics** | |
| $d_{32}$ | Sauter mean diameter (SMD) |
| $d_{\max}, D_{\max}$ | Maximum bubble diameter |
| $d_{99.8}$ | $99.8^{\text{th}}$ percentile bubble diameter |
| $d_H, d_{\text{Hinze}}$ | Hinze critical diameter for turbulent breakup |
| $A$ | Projected 2D bubble area |
| $n_i, N$ | Number of bubbles of diameter |
| **Bubble size distribution** | |
| $f(d)$ | Modified Gaussian Distribution function of bubble diameter |
| $\Phi$ | Cumulative bubble size distribution |
| $d_g$ | Geometric mean bubble diameter |
| $\sigma_g$ | Geometric standard deviation of bubble diameter |
| **Dimensionless and empirical quantities** | |
| $\text{Re}, \text{Re}_\tau, \text{Re}_\lambda$ | Reynolds numbers (bulk, friction, and Taylor-scale) |
| $We_{\text{crit}}$ | Critical Weber number for breakup onset |
| $\kappa, B$ | von Kármán constant; log-law constant |
| $\Gamma, h$ | Coalescence rate and Collision frequency |
| $\tau, T$ | Time period; averaging period |
| $\Pi$ | Empirical factor |

Table 4. List of symbols and their descriptions used in the manuscript.

## Appendix A. Turbulent kinetic energy ($tke$, near wall)

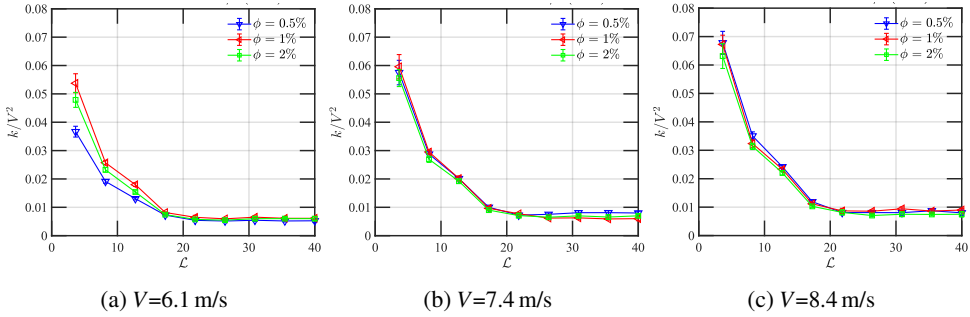

(a) $V$ =6.1 m/s      (b) $V$ =7.4 m/s      (c) $V$ =8.4 m/s

Figure 16. Axial variation of $tke$ ($k$) near wall plane for three bulk velocities $V(Q)$ and three void fractions $\phi$.

## Appendix B. Turbulent dissipation ($\varepsilon$, near wall)

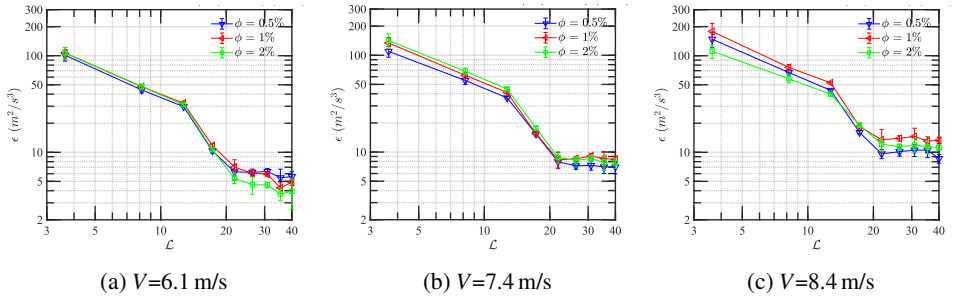

(a) $V$=6.1 m/s      (b) $V$=7.4 m/s      (c) $V$=8.4 m/s

Figure 17. Axial variation of $\varepsilon$ near wall plane for three bulk velocities $V(Q)$ and three void fractions $\phi$.

## Appendix C. Buble size growth in coalescence regime

### C.1. *Turbulent collision kernel*

Effective rate of coalescence [= collision frequency ($h$) × coalescence efficiency ($\lambda$)] can be written as:

$$\Gamma = h\,\lambda$$

The classical scaling (Kolmogorov 1991), originating with Kolmogorov and presented in standard treatments (Batchelor 1953; Monin & Yaglom 1975), underpins inertial–range collision models for bubbly dispersions. In the inertial subrange of turbulence, the velocity increment across a bubble scale $d$ follows Kolmogorov's similarity hypothesis, which results in (Batchelor 1953; Monin & Yaglom 1975)

$$u_{\text{rel}}(d) \sim (\varepsilon\,d)^{1/3},$$

where $\varepsilon$ is the mean dissipation rate. Multiplying the inertial–range relative speed by the geometric cross–section $\sim d^2$ gives the turbulent collision kernel

$$\beta \propto u_{\text{rel}}(d) \times d^2 \propto \varepsilon^{1/3}\,d^{7/3},$$

a form widely used in population–balance closures for bubble–bubble encounters in turbulent flows (Saffman & Turner 1956; Prince & Blanch 1990a; Luo & Svendsen 1996). Further, Prince & Blanch (1990b) and Luo & Svendsen (1996) modeled the collision frequency for two bubbles of diameters $d_i$ and $d_j$ in isotropic turbulence as,

$$h = C\,(\varepsilon)^{1/3}\,(d_i + d_j)^2 \cdot \sqrt{d_i^{\frac{2}{3}} + d_j^{\frac{2}{3}}} \qquad \text{(C 1)}$$

where $C$ is a constant. For a simplified scaling law, assuming a monodisperse system ($d_i = d_j = d$) results in $(d_i + d_j)^3 = (2d)^3 = 8d^3$, and $\Gamma$ simplifies to

$$\Gamma = 8C\,\lambda\,(\varepsilon)^{1/3}\,d^{\frac{7}{3}} \tag{C2}$$

with units of volume per time, and $\lambda \in [0, 1]$.

## C.2. *Population balance → diameter growth*

The number density $n$ can be written using Smoluchowski's equation for binary coagulation (avoiding double counting and assuming monodisperse conditions) (Kreer & Penrose 1994):

$$\frac{dn}{dt} = -\frac{1}{2}\,\Gamma\,n^2. \tag{C3}$$

For a duct flow with mean velocity $V$, use $t = \mathcal{L}/V$. Air-volume conservation relates $n$ and $d$:

$$\phi = n\,\frac{\pi d^3}{6} \qquad \Rightarrow \qquad n = \frac{6\phi}{\pi d^3}. \tag{C4}$$

Eliminating $n$ yields the identity (valid for any kernel):

$$\frac{dn}{dt} = \frac{dn}{dd}\frac{dd}{dt} = \frac{d}{dd}\left(\frac{6\phi}{\pi d^3}\right)\frac{dd}{dt} = -\frac{18\phi}{\pi d^4}\frac{dd}{dt}. \tag{C5}$$

After simplification,

$$\frac{dd}{d\mathcal{L}} = \frac{\Gamma\,\phi}{\pi\,d^2}. \tag{C6}$$

Inserting $\Gamma = 8C\,\lambda(\varepsilon)^{1/3}d^{\frac{7}{3}}$ gives

$$\frac{dd}{d\mathcal{L}} = \frac{8C}{\pi}\,\lambda\,\phi\,(\varepsilon)^{1/3}\,d^{\frac{1}{3}} \tag{C7}$$

Bubble diameter growth law with decaying turbulence ($\varepsilon \to \varepsilon(\mathcal{L})$):

$$\frac{dd}{d\mathcal{L}} = K[\varepsilon(\mathcal{L})]^{1/3}d^{\frac{1}{3}}, \qquad \text{where } K = \frac{8C}{\pi}, \lambda, \phi. \tag{C8}$$

## C.3. *Simplifying coalescence efficiency ($\lambda$)*

The coalescence efficiency can be described by the widely known film–drainage (FD) model,

$$\lambda_{ij} = \exp\left(-\frac{t_{\text{drain}}}{t_{\text{contact}}}\right), \tag{C9}$$

where $t_{\text{drain}}$ is the time for the intervening liquid film to thin to rupture thickness and $t_{\text{contact}}$ is the hydrodynamic contact time during a collision (Lehr et al. 2002a). In bubbly turbulence with deformable, mobile interfaces, FD formulations distinguish regimes by interfacial mobility; for gas bubbles in clean liquids, the relevant asymptote is the inertia-controlled drainage limit (Chesters 1991; Chan et al. 2011; Liao & Lucas 2010a).

Using Chesters (1991) parallel–film model in the inertia limit, the drainage time reduces to

$$t_{\text{drain}} = \frac{1}{2}\,\frac{\rho_c\,u_{rel}\,r^2}{\sigma}, \qquad r = \frac{d}{2}, \tag{C10}$$

with continuous–phase density $\rho_c$, surface tension $\sigma$, bubble radius $r$, and approach speed $u_{rel}$ (Liao & Lucas 2010a).[1]

A standard turbulent contact time is

$$t_{\text{contact}} \sim \frac{d^{2/3}}{\varepsilon^{1/3}}, \tag{C11}$$

---

[1] For unequal sizes see the (Luo 1995) generalization; here we restrict to monodisperse.

and an inertial–range estimate for the approach velocity is

$$u_{rel} \approx \sqrt{2} \left[ \varepsilon (2d) \right]^{1/3}. \tag{C 12}$$

With monodisperse simplification ($d_i = d_j = d$), and combining (C 10)–(C 12) gives

$$\frac{t_{\text{drain}}}{t_{\text{contact}}} = \frac{\rho_c}{\sigma} 2^{-13/6} \varepsilon^{2/3} d^{5/3}, \tag{C 13}$$

$$\Rightarrow \quad \lambda(d) = \exp \left[ - A \frac{\rho_c}{\sigma} \varepsilon^{2/3} d^{5/3} \right], \qquad A = 2^{-13/6} \approx 0.223. \tag{C 14}$$

Using $\rho_c = 998 \, \text{kg m}^{-3}$ and $\sigma = 0.072 \, \text{N m}^{-1}$ (air–water), Equation (C 14) is fully specified.

- **Case A (72 LPM and 0.5%) at $\mathcal{L} = 0$:** $\varepsilon = 200 \, \text{m}^2 \, \text{s}^{-3}$, $d = 200 \, \mu\text{m}$. From (C 12) $u_{rel} \approx 0.609 \, \text{m s}^{-1}$; using (C 10)–(C 11), $t_{\text{drain}} \approx 4.22 \times 10^{-5}$ s, $t_{\text{contact}} \approx 5.85 \times 10^{-4}$ s, so

$$\lambda = \exp \left( -\frac{t_{\text{drain}}}{t_{\text{contact}}} \right) \approx \exp(-0.072) \approx \mathbf{0.93}.$$

- **Case B (72 LPM and 0.5%) at $\mathcal{L} = 40$:** $\varepsilon = 10 \, \text{m}^2 \, \text{s}^{-3}$, $d = 500 \, \mu\text{m}$. $u_{rel} \approx 0.305 \, \text{m s}^{-1}$, $t_{\text{drain}} \approx 1.32 \times 10^{-4}$ s, $t_{\text{contact}} \approx 2.92 \times 10^{-3}$ s, hence

$$\lambda \approx \exp(-0.045) \approx \mathbf{0.95}.$$

Hence, it can be concluded that $\lambda$ remains high and constant for our scaling regime.

$$\lambda \sim \text{constant}, \quad \Rightarrow \quad K \sim \text{constant} \tag{2}$$

## C.4. *Solving the population balance equation*

Using the decay as a power law referenced to a finite start time $\mathcal{L} > 0$ (refer Equation 4.1):

$$\varepsilon(t) = \varepsilon_0 \left( \mathcal{L} \right)^{-m}, \qquad \varepsilon_0 \equiv \varepsilon(\mathcal{L}), \tag{C 15}$$

then simplifying Eq. C 7

$$\frac{1}{d^{1/3}} \frac{\mathrm{d}d}{\mathrm{d}\mathcal{L}} = K (\varepsilon_0)^{1/3} \mathcal{L}^{-m/3}. \tag{C 16}$$

$$d(\mathcal{L}) = \left[ d_0^{2/3} + K (\varepsilon_0)^{1/3} \int_{\mathcal{L}_0}^{\mathcal{L}} \mathcal{L}^{-m/3} \, d\mathcal{L} \right]^{3/2} \tag{2}$$

$$d(\mathcal{L}) = \left[ d_0^{2/3} + K \varepsilon_0^{1/3} \mathcal{L}_0^{1/3} \left( \mathcal{L}^{\frac{3-m}{3}} - \mathcal{L}_0^{\frac{3-m}{3}} \right) \right]^{3/2}$$

$$d(\mathcal{L}) \propto \mathcal{L}^{\frac{3-m}{2}} \propto \mathcal{L}^{\beta} \text{ here } \beta = \frac{3-m}{2}$$

*Case A: $\varepsilon \propto \mathcal{L}^{-1.8}$ ($m = 1.8$)*

$$\int_{\mathcal{L}_0}^{\mathcal{L}} \mathcal{L}^{-0.6} d\mathcal{L} = \frac{5}{2} \left( \mathcal{L}^{0.4} - \mathcal{L}_0^{0.4} \right),$$

$$d(\mathcal{L}) = \left[ d_0^{2/3} + K \varepsilon_0^{1/3} \mathcal{L}_0^{1/3} \left( \mathcal{L}^{0.4} - \mathcal{L}_0^{0.4} \right) \right]^{3/2}$$

$$\boxed{d(\mathcal{L}) \propto \mathcal{L}^{0.6}}$$

Table 5. Bubble diameter scaling in highly decaying turbulent flow

| $\varepsilon$ scaling | $d(d_{32}$ and $d_{99.8})$ scaling |
|---|---|
| $\varepsilon \propto \mathcal{L}^{-m}$ | $d \propto \mathcal{L}^{\beta}, \ \beta = \dfrac{3-m}{2}$ |
| $\varepsilon \propto \mathcal{L}^{-1.8}$ | $d \propto \mathcal{L}^{3/5}$ |
| $\varepsilon \propto \mathcal{L}^{-2}$ | $d \propto \mathcal{L}^{1/2}$ |
| $\varepsilon \propto \mathcal{L}^{-2.2}$ | $d \propto \mathcal{L}^{2/5}$ |

*Case B: $\varepsilon \propto \mathcal{L}^{-2}$ (m = 2)*

$$\int_{\mathcal{L}_0}^{\mathcal{L}} \mathcal{L}^{-0.67} d\mathcal{L} = \frac{10}{3} \left( \mathcal{L}^{0.33} - \mathcal{L}_0^{0.33} \right),$$

$$d(\mathcal{L}) = \left[ d_0^{2/3} + K \, \varepsilon_0^{1/3} \mathcal{L}_0^{1/3} \left( \mathcal{L}^{0.33} - \mathcal{L}_0^{0.33} \right) \right]^{3/2}$$

$$\boxed{d(\mathcal{L}) \propto \mathcal{L}^{0.5}}$$

*Case C: $\varepsilon \propto \mathcal{L}^{-2.2}$ (m = 2.2)*

$$\int_{\mathcal{L}_0}^{\mathcal{L}} \mathcal{L}^{-0.73} d\mathcal{L} = \frac{10}{2.7} \left( \mathcal{L}^{0.27} - \mathcal{L}_0^{0.27} \right),$$

$$d(\mathcal{L}) = \left[ d_0^{2/3} + K \, \varepsilon_0^{1/3} \mathcal{L}_0^{1/3} \left( \mathcal{L}^{0.27} - \mathcal{L}_0^{0.27} \right) \right]^{3/2} \qquad \boxed{d(\mathcal{L}) \propto \mathcal{L}^{0.4}}$$